\documentclass[a4paper,fleqn,10pt]{article}
\pdfoutput=1
\usepackage{amsmath}
\usepackage{amssymb}
\usepackage{array}
\usepackage{calc}
\usepackage{longtable}
\usepackage{multirow}
\usepackage{graphicx}
\usepackage{xspace}

\numberwithin{equation}{section}
\usepackage{cite}
\usepackage[a4paper,pdfborder={0 0 0}]{hyperref}
\usepackage[format=hang,labelfont=bf,hypcap=true]{caption}
\usepackage{subfig}
\usepackage{sectsty}
\allsectionsfont{\sffamily}
\subsubsectionfont{\mdseries\itshape\large}
\makeatletter
\DeclareRobustCommand*{\bfseries}{%
  \not@math@alphabet\bfseries\mathbf
  \fontseries\bfdefault\selectfont
  \boldmath
}
\makeatother
\let\spreprint\empty
\newcommand{\preprint}[1]{\def\spreprint{\protect#1}}
\let\sinstitute\empty
\newcommand{\institute}[1]{\def\sinstitute{\protect#1}}
\makeatletter
\renewcommand{\maketitle}{\begingroup
  \null\thispagestyle{empty}%
    \ifx\spreprint\empty
      \vskip 5ex
    \else
      \flushright\large\spreprint\vskip 2ex
    \fi
    \vskip 5ex
    \flushleft
      {\sffamily\bfseries\scalebox{1.0}{
       \begin{minipage}{\textwidth}\centering\huge\@title\end{minipage}
      }}\vskip 6ex
      \@author\vskip 2ex
      \ifx\sinstitute\empty
      \else
        {\small\sinstitute}
      \fi
    \vskip 5ex
  \endgroup
}
\makeatother
\renewenvironment{abstract}{\begin{center}
  {\large\sffamily\bfseries Abstract: }
  \begin{minipage}[t]{0.75\textwidth}
}{\end{minipage}\end{center}\vskip 10ex}

\numberwithin{equation}{section}
\newcommand{\MCatNLO}{M\protect\scalebox{0.8}{C}@N\protect\scalebox{0.8}{LO}\xspace}

\newcommand{\POWHEG}{P\protect\scalebox{0.8}{OWHEG}\xspace}

\newcommand{\MEPS}{M\scalebox{0.8}{E}P\scalebox{0.8}{S}\xspace}

\newcommand{\MENLOPS}{ME\protect\scalebox{0.8}{NLO}PS\xspace}
\newcommand{\MEPSatNLO}{M\scalebox{0.8}{E}P\scalebox{0.8}{S}@N\protect\scalebox{0.8}{LO}\xspace}

\newcommand{\Blackhat}{B\protect\scalebox{0.8}{LACK}H\protect\scalebox{0.8}{AT}\xspace}

\newcommand{\Sherpa}{S\protect\scalebox{0.8}{HERPA}\xspace}
\newcommand{\Comix}{C\protect\scalebox{0.8}{OMIX}\xspace}

\newcommand{\Amegic}{A\protect\scalebox{0.8}{MEGIC++}\xspace}

\long\def\symbolfootnote[#1]#2{\begingroup%
\def\thefootnote{\fnsymbol{footnote}}\footnote[#1]{#2}\endgroup}
\newcommand{\EqRef}[1]{Eq.~\eqref{#1}}

\newcommand{\abr}[1]{\langle #1\rangle}
\newcommand{\cbr}[1]{\left\{ #1\right\}}
\newcommand{\sbr}[1]{\left[ #1\right]}

\newcommand{\done}{{\rm d}}
\newcommand{\order}{\mathcal{O}}
\newcommand{\mc}[1]{\mathcal{#1}}
\newcommand{\mr}[1]{\mathrm{#1}}
\newcommand{\mb}[1]{\mathbb{#1}}

\newcommand{\bea}{\begin{eqnarray}}
\newcommand{\eea}{\end{eqnarray}}
\newcommand{\bi}{\begin{itemize}}
\newcommand{\ei}{\end{itemize}}

\hypersetup{
  pdfauthor={Thomas Gehrmann, Stefan Hoeche, Frank Krauss,
    Marek Schoenherr, Frank Siegert},
  pdftitle={NLO QCD matrix elements + parton showers in ee -> hadrons}
}
\preprint{ZU-TH 11/12\\SLAC-PUB 15190\\
  IPPP/12/51\\DCPT/12/102\\LPN12-080\\
  FR-PHENO-2012-018\\MCNET-12-08}
\author{Thomas Gehrmann$^1$, Stefan H{\"o}che$^2$, Frank Krauss$^3$,
  Marek Sch{\"o}nherr$^3$, Frank Siegert$^4$}
\title{NLO QCD matrix elements + parton showers\\ 
  in $e^+e^-\to$ hadrons}
\institute{$^1$ Institut f{\"u}r Theoretische Physik, 
  Universit{\"a}t Z{\"u}rich, CH-8057 Z{\"u}rich, Switzerland\\
  $^2$ SLAC National Accelerator Laboratory, 
  Menlo Park, CA 94025, USA\\
  $^3$ Institute for Particle Physics Phenomenology,
  Durham University, Durham DH1 3LE, UK\\
  $^4$ Physikalisches Institut, Albert-Ludwigs-Universit{\"a}t Freiburg,
  Hermann-Herder-Str. 3, D-79104 Freiburg, Germany\\}
\begin{document}
\maketitle
\begin{abstract}
  We present a new approach to combine multiple NLO parton-level calculations 
matched to parton showers into a single inclusive event sample. The method 
provides a description of hard multi-jet configurations at next-to leading 
order in the perturbative expansion of QCD, and it is supplemented with the 
all-orders resummed modelling of jet fragmentation provided by the parton 
shower.  The formal accuracy of this technique is discussed 
in detail, invoking the example of electron-positron annihilation into hadrons.
We focus on the effect of renormalisation scale variations in particular.  
Comparison with experimental data from LEP underlines that this novel 
formalism describes data with a theoretical accuracy that has hitherto 
not been achieved in standard Monte Carlo event generators.

\end{abstract}
\section{Introduction}

During the past decade, Monte-Carlo methods for simulating hadronic final states 
in collider experiments have improved continuously. Multi-purpose event generators
incorporating the most recent higher-order perturbative QCD calculations 
have thus emerged, making them available to phenomenology and experiment alike. 
This has far-reaching consequences for both precision physics and searches for new phenomena.  
Key to the developments has been the steady progress in understanding the interplay 
of real and virtual higher-order QCD corrections on one hand and of resummation 
techniques like parton-shower algorithms on the other hand.  The construction and
development of simulation tools for QCD processes has become one of the central 
activities of research in collider phenomenology.

This publication discusses an extension to the established techniques
of multi-jet merging and next-to-leading order matrix-element matching.
Existing multi-jet merging methods (\MEPS) combine leading-order 
matrix elements of varying final-state multiplicity with parton showers.
They were pioneered in~\cite{Catani:2001cc,Lonnblad:2001iq,Mangano:2001xp,Krauss:2002up}
and further matured in~\cite{Hoeche:2009rj,Hamilton:2009ne,Giele:2011cb,Lonnblad:2011xx}.
The key advantage of these methods is the possibility to describe arbitrarily
complex final states at leading order in the strong coupling, while providing
fully inclusive event samples with resummation effects taken into account. 
They have therefore become standard analysis tools for collider experiments.
However, they lack the precision of a full next-to-leading order perturbative
calculation. This is remedied by next-to-leading order matrix-element matching 
methods (\MCatNLO), which combine NLO QCD calculations of fixed jet multiplicity 
with parton showers. They were introduced in~\cite{Frixione:2002ik,Nason:2004rx} 
and have recently been automated in various programs~\cite{Frederix:2011ig,Hoeche:2011fd}. 
Their main advantage lies in the excellent description of well-defined, inclusive 
final states. Using the \MENLOPS technique~\cite{Hamilton:2010wh,Hoeche:2010kg}, 
it is possible to make these results exclusive and combine them with higher-multiplicity leading-order 
predictions in order to recover the virtues of \MEPS methods.

The aim of this paper is to further improve upon the existing algorithms and construct 
a consistent, process-independent merging method for matched NLO predictions 
with varying jet multiplicity. Pictorially speaking, we intend to replace 
the leading-order matrix elements of the original \MEPS approach with corresponding 
ones at next-to-leading order. This is achieved by combining various \MCatNLO 
event samples and accounting for potential double counting by means of a modified 
truncated parton shower~\cite{Nason:2004rx,Hoeche:2009rj}. Ultimately, we intend to maintain the fixed-order accuracy 
of the matrix elements, but also to preserve the logarithmic accuracy of the parton shower.  
The new method discussed here goes well beyond the scope of the \MENLOPS technique.

In the framework of this paper the formalism is specified for a multi-jet 
merging at NLO accuracy for $e^+e^-$-annihilations into hadrons, building 
on the existing implementations of \MEPS~\cite{Hoeche:2009rj} and 
\MCatNLO~\cite{Hoeche:2011fd} techniques in the \Sherpa event 
generator~\cite{Gleisberg:2003xi,Gleisberg:2008ta}.  In the present paper,
however, we will assume that the evolution parameter of the parton shower
is defined in such a way, equivalent to the measure of hardness of a parton
splitting, that effects due to a mismatch of these two quantities can be
neglected.  In other words we will neglect effects that arise from allowed emissions 
generated by truncated parton showers.  An algorithm with the same 
goals and a similar setup for the parton shower has been detailed, also for 
$e^+e^-$-annihilations into hadrons, in~\cite{Lavesson:2008ah}.
A method for merging NLO vector boson plus 0 and 1-jet samples was introduced 
in~\cite{Alioli:2011nr}, while~\cite{Hoeche:2012yf} proposed a general method
for NLO vector boson production plus $n$ jets and implemented it for n=0,1,2.
Here we apply the method of~\cite{Hoeche:2012yf} to hadronic final states 
in $e^+e^-$-annihilation.

The outline of the present paper is as follows: Section~\ref{Sec::KnownAlgos} 
discusses the \MEPS algorithm for matrix-element merging at leading order, 
and the \MCatNLO method for NLO matching as implemented in \Sherpa.  
As an intermediate step, the implementation of the \MENLOPS idea for
\MCatNLO core processes is presented.  With the notation
thus established, the new merging method at next-to leading order, \MEPSatNLO,
is introduced in Sec.~\ref{Sec:NLOMerging}.  The renormalisation scale
dependence of the result is discussed in some detail.  
Sec.~\ref{sec:mc-implementation} is devoted to details of the Monte-Carlo 
implementation.
Example results for the case of electron-positron annihilation into hadrons
are shown in Sec.~\ref{Sec:Results}, including the impact of scale variations
and of varying the number of jets described by NLO matrix element calculations.  
Sec.~\ref{Sec:Conclusions} presents our conclusions.

\section{Brief review of merging and matching techniques}
\label{Sec::KnownAlgos}
In this section, existing matrix-element parton-shower merging and matching 
methods are briefly reviewed, using the notation of~\cite{Hoeche:2011fd}.
As already stated in the introduction, the effects of allowed emissions 
generated by truncated showers~\cite{Nason:2004rx,Hoeche:2009rj} are ignored, which improves the readability
of this publication, allowing to focus on the structure of the result.  
For a full algorithmic solution, we refer to the parallel publication, in~\cite{Hoeche:2012yf}.
Our approach is justified by the choice of transverse momentum as evolution variable 
in the parton shower used for this publication.

In the context of merging, we define a jet criterion $Q_n$, 
which typically denotes the minimal value of some relative transverse momentum 
present in the $n$-parton phase-space configuration $\Phi_n$.  Correspondingly, 
$Q_{\rm cut}$ denotes a jet-defining cut value, called the merging scale, such 
that for $n$-jet events the condition $Q_n>Q_{\rm cut}$ is fulfilled\footnote{
  The jet criterion $Q$ applied here has been slightly modified compared 
  to~\cite{Hoeche:2009rj}, in order to reflect the fact that no unique parton
  flavour can be assigned at the next-to-leading order. For any pair of 
  final-state partons $i$ and $j$ we define
  \begin{equation}\label{eq:jet_criterion}
    Q_{ij}^2\,=\;2\,p_i p_j\,\min\limits_{k\ne i,j}\,
    \frac{2}{C_{i,j}^k+C_{j,i}^k}\;
    \qquad\text{where}\qquad
    C_{i,j}^k\,=\;\frac{p_ip_k}{(p_i+p_k)p_j}\;.
  \end{equation}
  The spectator index $k$ runs over all possible coloured particles.}.

Formally, the quantity of interest is the expectation value $\abr{O}$ of an 
arbitrary, infrared-safe observable $O$, evaluated by taking the average
over sufficiently many points in an $n$-particle phase-space, $\Phi_n$.

The methods reviewed here, as well as our newly proposed technique, 
have the following aims
\begin{itemize}
\item Multi-jet merging techniques\\
  For configurations with $n$ jets, the respective fixed-order
  accuracy of $\abr{O}$ inherent to the parton-level result should be maintained.  
  More precisely, for leading-order merging (\MEPS), jet observables for $n$ 
  jets above the merging scale $Q_{\rm cut}$ should be determined at 
  leading-order accuracy. For next-to-leading order merging (\MEPSatNLO)
  they should be given at NLO accuracy. For configurations below $Q_{\rm cut}$,
  the \MEPS accuracy will be that of the parton shower, while for \MEPSatNLO 
  leading-order accuracy is envisaged.
  At the same time we require that the logarithmic accuracy of the shower be
  either maintained or improved in the region above $Q_{\rm cut}$.
\item NLO matching methods\\
  For processes leading to $n$-parton final states at leading order all 
  $n$-particle inclusive observables, and in particular the total cross section,
  are expected to reproduce the fixed order NLO results.  
  At the same time, all $n+1$-particle observables are expected to be given
  at leading order accuracy, while higher-order emissions should still be
  described by the leading logarithmic approximation of the parton shower.
\end{itemize}

\subsection{Leading-order merging - \protect\MEPS}
\label{Sec:lomerging}
In the context of the leading-order merging method proposed in~\cite{Hoeche:2009rj},
the following quantities are introduced:
\begin{itemize}
\item Squared leading-order (Born) matrix elements, ${\rm B}_n(\Phi_n)$, 
  for $n$ outgoing particles, summed (averaged) over final state 
  (initial state) spins and colours and including symmetry and flux factors.
\item Sudakov form factors of the parton shower, given by 
  \begin{equation}
    \Delta_n^\text{(PS)}(t,t') = 
    \exp\Bigg\{-\int\limits_t^{t'}\done\Phi_1\;\mr{K}_n(\Phi_1)\Bigg\}\,,
  \end{equation}
  $\mr{K}_n$ denotes the sum of all splitting kernels for the $n$-body 
  final state. The one-particle phase-space element for a splitting, 
  $\done\Phi_1$, is parametrised as
  \begin{equation}\label{eq:one-dim-ps}
    \done\Phi_1\,=\;\done t\,\done z\,\done\phi\;\,J(t,z,\phi)\;, 
  \end{equation}
  where $t$ is the ordering variable, $z$ is the splitting variable 
  and $\phi$ is the azimuthal angle. $J(t,z,\phi)$ is the appropriate 
  Jacobian factor. The ordering variable is usually taken to fulfil 
  $t\propto k_\perp^2$ as $t\to 0$. 
\item The resummation scale $\mu_Q$, which defines an upper limit 
  of parton evolution in terms of the shower evolution variable. 
  $t_c$ is an infrared regulator of the order of $\Lambda_\text{QCD}$ 
  marking the transition into the non-perturbative region.
\end{itemize}
The expectation value of an arbitrary, infrared-finite observable $O$,
leading order for $n$ partons, to $\order(\alpha_s)$
has been computed in~\cite{Hoeche:2010kg}. It is derived from the following expression:
\begin{equation}\label{eq:mepslo}
  \begin{split}
    \abr{O}\,=&\;
    \int\done\Phi_n\; {\rm B}_n(\Phi_n)
     \Bigg[\;
	\Delta_n^\text{(PS)}(t_c,\mu_Q^2)\,O(\Phi_n)
	+\int\limits_{t_c}^{\mu_Q^2}\done\Phi_1\; {\rm K}_n(\Phi_1)\,
	 \Delta_n^\text{(PS)}(t_{n+1},\mu_Q^2)\,\Theta(Q_\text{cut}-Q_{n+1})\;O(\Phi_{n+1})
    \;\Bigg]\\
    &\;+\int\done\Phi_{n+1}\;{\rm B_{n+1}}(\Phi_{n+1})\,
      \Delta_n^\text{(PS)}(t_{n+1},\mu_Q^2)\,\Theta(Q_{n+1}-Q_\text{cut})\;O(\Phi_{n+1})\;,
  \end{split}
\end{equation}
where $O(\Phi_{m})$ is the observable evaluated for an $m$-parton final state.
The square bracket on the first line and the Sudakov factor on the second line are
both generated by the parton shower, while the terms $\done\Phi_n \mr{B}_n$ and
$\done\Phi_{n+1} \mr{B}_{n+1}$ correspond to the fixed-order event generation.
The term on the second line yields leading-order accuracy for any $n+1$-particle observable 
in the region $Q_{n+1}>Q_{\rm cut}$. Leading-order accuracy for observables sensitive to $\Phi_n$ 
is guaranteed by the fact that \EqRef{eq:mepslo} can be rewritten as 
\begin{equation}\label{eq:mepslo_pscomp}
  \begin{split}
    \abr{O}\,=&\;\int\done\Phi_n\; {\rm B}_n(\Phi_n)
     \Bigg[\;
	\Delta_n^\text{(PS)}(t_c,\mu_Q^2)\,O(\Phi_n)
	+\int\limits_{t_c}^{\mu_Q^2}\done\Phi_1\; {\rm K}_n(\Phi_1)\,
	 \Delta_n^\text{(PS)}(t_{n+1},\mu_Q^2)\;O(\Phi_{n+1})
    \;\Bigg]\\
    &\;+\int\done\Phi_{n+1}\;\Big[\,{\rm B_{n+1}}(\Phi_{n+1})-
       {\rm B_n}(\Phi_{n})\,\mr{K}_n(\Phi_{n+1})\,\Big]\,
      \Delta_n^\text{(PS)}(t_{n+1},\mu_Q^2)\,\Theta(Q_{n+1}-Q_\text{cut})\;O(\Phi_{n+1})\;,
  \end{split}
\end{equation}
where the first line is the $\mc{O}(\alpha_s)$ parton-shower result~\cite{Hoeche:2010pf} 
and independent of $Q_{\rm cut}$.
The additional terms on the second line incorporate possible sub-leading colour
single logarithms as well as power corrections.  The size of these corrections determines the 
potential discontinuity in $\abr{O}$ at $Q_{\rm cut}$. It can be large if $Q_{\rm cut}$ 
is either far from the collinear limit or sub-leading colour single logarithms 
are important. Sub-leading colour configurations, however, can be included in a 
systematic manner, as was detailed in~\cite{Hoeche:2011fd}.

An important feature of Eq.~\eqref{eq:mepslo} is that it can be iterated to incorporate
higher-multiplicity leading-order matrix elements into the prediction. By replacing
$n\to n+1$, all properties of the algorithm remain the same. In order to obtain this
property when dealing with next-to-leading order matrix elements, a slight modification 
is necessary, which will be described in Sec.~\ref{Sec:NLOMerging}.

\subsection{Next-to-leading order matching - \protect\MCatNLO}
\label{Sec:matching}
In the \MCatNLO matching method the following additional quantities are 
needed:
\begin{itemize}
\item Squared real-emission matrix elements, ${\rm R}_n(\Phi_{n+1})$,
  for $n$-particle processes, summed (averaged) over final state 
  (initial state) spins and colours and including symmetry and flux factors.
  Note that ${\rm R}_n(\Phi_{n+1})={\rm B}_{n+1}(\Phi_{n+1})$.
\item The NLO-weighted Born differential cross section $\bar{\rm B}_n^{\rm (A)}$,
  defined as
  \begin{equation}\label{eq:mcatnlo_bbar}
    \begin{split}
      \bar{\rm B}_n^{\rm (A)}(\Phi_n)\,=&\;
          {\rm B}_n(\Phi_n)+\mr{V}_n(\Phi_n)+{\rm I}_n^{\rm (S)}(\Phi_n)
          \vphantom{\int}\\
          &\;{}
          +\int\done\Phi_1
          \sbr{\,
            {\rm D}_n^{\rm (A)}(\Phi_{n+1})\,\Theta(\mu_Q^2-t_{n+1})
            -{\rm D}_n^{\rm (S)}(\Phi_{n+1})
          \,}\;.
    \end{split}
  \end{equation}
  Here, ${\rm V}_n$ is the Born-contracted one-loop amplitude, 
  and ${\rm I}_n^{\rm (S)}$ is the sum of integrated subtraction terms, 
  cf.~\cite{Hoeche:2011fd}, while ${\rm D}_n^{\rm (S)}$ are the 
  corresponding real subtraction terms.  In contrast, ${\rm D}_n^{\rm (A)}$ 
  are the \MCatNLO evolution kernels multiplied by Born matrix elements.
  Both functions can be decomposed in terms of dipole 
  contributions, ${\rm D}=\sum_{ij,k}{\rm D}_{ij,k}$, where each dipole 
  encodes exactly one singular region~\cite{Hoeche:2011fd}.  Further,  each 
  dipole has a corresponding phase space factorisation $\done\Phi_{n+1}=
  \done\Phi_n\,\done\Phi_1^{ij,k}$ and $t_{n+1} = t(\Phi_{n+1})$ is defined in 
  terms of \EqRef{eq:one-dim-ps} in each of these dipole phase spaces.
\item The hard remainder function
  \begin{equation}\label{eq:mcatnlo_h}
    \mr{H}_n^{\rm(A)}(\Phi_{n+1})\;=\;
    \mr{R}_n(\Phi_{n+1})-\mr{D}_n^{\rm(A)}(\Phi_{n+1})\,
    \Theta(\mu_Q^2-t_{n+1})\,,
  \end{equation}
  with $t_{N+1} = t(\Phi_{n+1})$ defined as above.
\item The \MCatNLO Sudakov form factor
  \begin{equation}\label{def_mcatnlosud}
    \begin{split}
      \Delta_n^\text{(A)}(t,t')
      \,=&\;
      \exp\Bigg\{-\int\limits_t^{t'}\done\Phi_1\;
      \frac{{\rm D}_n^{\rm (A)}(\Phi_n,\Phi_1)}{{\rm B}(\Phi_n)}\Bigg\}\;,
    \end{split}
  \end{equation}
  Note that $\Delta_n^{\rm(A)}$ implicitly depends on $\Phi_n$, while the 
  original Sudakov form factor $\Delta_n^{\rm(PS)}$ does not.  This is a 
  consequence of the fact that the two Sudakov form factors differ by their 
  treatment of colour and spin correlations and it was discussed in detail
  in~\cite{Hoeche:2011fd}. By incorporating full colour information in 
  $\mr{D}^{\rm(A)}$, it is easily possible to obtain the exact same 
  singularity structure as in the real-emission matrix 
  element~\cite{Catani:1996vz,Catani:2000ef}.
\end{itemize}
The expectation value of an arbitrary infrared safe observable $O$ 
to $\mc{O}(\alpha_s)$ is then given by~\cite{Frixione:2002ik}
\begin{equation}\label{eq:mcatnlo}
  \begin{split}
    \abr{O}\,=&\;
    \int\done\Phi_n\; \bar{\rm B}_n^\text{(A)}(\Phi_n)
    \Bigg[\;
	\Delta_n^\text{(A)}(t_c,\mu_Q^2)\,O(\Phi_n)
	+\int\limits_{t_c}^{\mu_Q^2}\done\Phi_1\;
	 \frac{{\rm D}_n^\text{(A)}(\Phi_n,\Phi_1)}{{\rm B}_n(\Phi_n)}\,
	 \Delta_n^\text{(A)}(t_{n+1},\mu_Q^2)\;O(\Phi_{n+1})
    \;\Bigg]\\
    &\;{}
    +\int\done\Phi_{n+1}\;\mr{H}_n^{\rm(A)}(\Phi_{n+1})\,\;O(\Phi_{n+1})\;.
  \end{split}
\end{equation}
The square bracket on the first line is generated by a weighted parton shower,
which will be discussed in Sec.~\ref{Sec:mc_color}, 
while the terms $\done\Phi_n \bar{\mr{B}}_n^{\rm(A)}$ and
 $\done\Phi_{n+1} \mr{H}_{n}^{\rm(A)}$ correspond to fixed-order events.
Events generated according to the first line are referred to as standard, or $\mb{S}$-events,
while events generated according to the second line, the hard remainder, correspondingly 
are dubbed $\mb{H}$-events~\cite{Frixione:2002ik,Hoeche:2011fd}.  Note 
that the square bracket in the first line integrates to one, reflecting the 
probabilistic nature of the Sudakov form factor.  This, together with
equations~\eqref{eq:mcatnlo_bbar} and~\eqref{eq:mcatnlo_h}, implies that the 
total cross section reproduces the exact NLO result.  Correspondingly, an 
\MCatNLO prediction is next-to-leading order accurate for observables 
sensitive to the Born phase-space variables $\Phi_n$, and leading-order 
accurate for observables sensitive to $\Phi_{n+1}$.  In contrast to 
the \MEPS method, leading-order accuracy is maintained throughout the 
$n+1$-particle phase space, but it cannot be extended to higher parton or jet 
multiplicity. This extension will be the topic of Sec.~\ref{Sec:menlops}.

\subsection{Combining NLO matching and LO merging - \protect\MENLOPS}
\label{Sec:menlops}
NLO-matched predictions as described in Sec.~\ref{Sec:matching} can easily 
be merged with higher-multiplicity event samples at leading order accuracy 
using  the techniques described in Sec.~\ref{Sec:lomerging}.  The original
algorithm, based on the \POWHEG method~\cite{Nason:2004rx,Frixione:2007vw},
was independently proposed in~\cite{Hamilton:2010wh} and~\cite{Hoeche:2010kg}.
In this publication we extend the method to \MCatNLO, which requires the 
introduction of the local $K$-factor
\begin{equation}\label{eq:mcatnlo_lkf}
  k_n^{\rm(A)}(\Phi_{n+1})\,=\;
	 \frac{\bar{\rm B}_n^\text{(A)}(\Phi_n)}{{\rm B}_n(\Phi_n)}
	 \left(
	       1-\frac{\mr{H}_n(\Phi_{n+1})}{{\rm R}_n(\Phi_{n+1})}
	 \right)+\frac{\mr{H}_n(\Phi_{n+1})}{{\rm R}_n(\Phi_{n+1})}\;.
\end{equation}
It is motivated by the behaviour of the underlying \MCatNLO event sample 
in terms of $\mb{S}$- and $\mb{H}$-events~\cite{Frixione:2002ik,Hoeche:2011fd}.
In the limit $\mr{H}_n^{\rm(A)}\to 0$, i.e.\ for configurations with a soft
additional parton, we obtain $k_n^{\rm(A)}(\Phi_{n+1})\to
\bar{\rm B}_n^\text{(A)}(\Phi_n)/{\rm B}_n(\Phi_n)$. In the limit 
$\mr{H}_n^{\rm(A)}\to \mr{R}_n^{\rm(A)}$, i.e.\ for configurations with a
hard additional parton, we have instead $k_n^{\rm(A)}(\Phi_{n+1})\to 1$.
Hence, the higher-multiplicity tree-level result is ``scaled up'' by
the local $K$-factor from \MCatNLO in the soft region, and it is 
left untouched in the hard region.
In both cases, however, the $n$-parton phase-space configuration in 
Eq.~\eqref{eq:mcatnlo_lkf} is determined by backward clustering, as described 
in~\cite{Hoeche:2009rj}.

The expectation value of an arbitrary, infrared-finite observable 
to $\mc{O}(\alpha_s)$ in the \MENLOPS method for \MCatNLO is given by
\begin{equation}\label{eq:menlops}
  \begin{split}
  \abr{O}\,=&\;
  \int\done\Phi_n\;\bar{\rm B}_n^\text{(A)}(\Phi_n)
  \\&\times
  \Bigg[\,
	\Delta_n^\text{(A)}(t_c,\mu_Q^2)\,O(\Phi_n)
	+\int\limits_{t_c}^{\mu_Q^2}\done\Phi_1\;
         \frac{{\rm D}_n^\text{(A)}(\Phi_n,\Phi_1)}{{\rm B}_n(\Phi_n)}\,
	 \Delta_n^\text{(A)}(t_{n+1},\mu_Q^2)\,\Theta(Q_\text{cut}-Q_{n+1})\;
         O(\Phi_{n+1})\;
  \,\Bigg]
  \\&\hspace*{-2mm}
  +\int\done\Phi_{n+1}\;\mr{H}_n^{\rm(A)}(\Phi_{n+1})\,
  \Delta_n^\text{(PS)}(t_{n+1},\mu_Q^2)\,\Theta(Q_\text{cut}-Q_{n+1})\;
   O(\Phi_{n+1})
   \\&\hspace*{-2mm}
  +\int\done\Phi_{n+1}\,k_n^{\rm(A)}(\Phi_{n+1})\,
   {\rm B }_{n+1}(\Phi_{n+1})\,\Delta_{n}^\text{(PS)}(t_{n+1},\mu_Q^2)\,
   \Theta(Q_{n+1}-Q_\text{cut})\;O(\Phi_{n+1})\;.
  \end{split}
\end{equation}
This prediction is next-to-leading order accurate for observables sensitive 
to $\Phi_n$ and leading-order accurate for observables sensitive to $\Phi_{n+1}$.
The key advantage compared to a pure NLO-matched prediction is that final 
states of higher jet multiplicity are treated as in the \MEPS approach.
The improvement over results obtained by \MEPS methods is the next-to leading 
order accuracy of the inclusive cross section and of observables sensitive to $\Phi_n$.

The method aims to maintain the full NLO-accuracy in the $n$-jet phase 
space and the LO-accuracy in the $(n+1)$-jet phase space, without upsetting
the logarithmic accuracy of the parton shower.  In order to see that this indeed
is the case, equation~\eqref{eq:menlops} can be rephrased as follows:
\begin{equation}\label{eq:split_menlops}
  \abr{O}\,=\;\abr{O}^{\rm MC@NLO}\,+\,\abr{O}^{\rm corr}\,,
\end{equation}
with $\abr{O}^{\rm MC@NLO}$ given by \eqref{eq:mcatnlo}, and thus showing the
desired property.  It thus remains to show that the correction term does
not introduce unwanted terms of higher logarithmic order.  Omitting the obvious phase
space arguments of the different matrix element contributions, it is given by
\begin{equation}\label{eq:menlops_hterm}
  \begin{split}
  \abr{O}^{\rm corr}\,=\,&\;
  \int\done\Phi_{n+1}\;\Theta(Q_{n+1}-Q_\text{cut})\,O(\Phi_{n+1})
  \,\Delta_n^{\rm(PS)}(t_{n+1},\mu_Q^2)  
  \\&\hspace*{10mm}\times\,
  \Bigg\{\,
  \Bigg[\frac{\bar{\mr{B}}_n^{\rm(A)}}{\mr{B}_n}
  \Bigg(1\,-\,\frac{\mr{H}^\text{(A)}_n}{\mr{B}_{n+1}}\Bigg)\,+\,
  \frac{\mr{H}^\text{(A)}_n}{\mr{B}_{n+1}}\Bigg]\,\mr{B}_{n+1}
  \,-\,\mr{H}^\text{(A)}_n\,-\,
  \frac{\bar{\mr{B}}_n^{\rm(A)}}{\mr{B}_n}
  \mr{D}_n^{\rm(A)}
  \frac{\Delta_n^{\rm(A)}(t_{n+1},\mu_Q^2)}{\Delta_n^{\rm(PS)}(t_{n+1},\mu_Q^2)}
  \Bigg\}
  \\ =\,&\;
  \int\done\Phi_{n+1}\;\Theta(Q_{n+1}-Q_\text{cut})\,O(\Phi_{n+1})\,
  \Delta_n^{\rm(PS)}(t_{n+1},\mu_Q^2)  
  \\&\hspace*{10mm}\times\, 
  \Bigg\{\,
  \frac{\bar{\mr{B}}_n^{\rm(A)}}{\mr{B}_n}
  \mr{D}_n^{\rm(A)}\Bigg(
  1\,-\,\frac{\Delta_n^{\rm(A)}(t_{n+1},\mu_Q^2)}{\Delta_n^{\rm(PS)}(t_{n+1},\mu_Q^2)}
  \Bigg)\Bigg\} 
  \end{split}
\end{equation}
Since $\mr{D}^\text{(A)}_n$ is of $\mc{O}(\alpha_sL^2)$ and because the ratio of
Sudakov form factor is at most of non-leading logarithmic order, 
$\mc{O}(\alpha_sL)$, and non-leading in $1/N_c$, the overall contribution of 
this term is at most of $\mc{O}(\alpha_s^2L^3)$.\footnote{This statement is based
  on the logarithmic accuracy of currently available parton showers. Parton showers
  which are extended to full NLL accuracy may become available in the future, 
  in which case the mismatch of $\mc{O}(\alpha_s^2L^3)$ would be absent.} 
The logarithmic accuracy of
the \MENLOPS method therefore depends entirely on the logarithmic accuracy 
of the parton shower. If the parton shower is correct to NLL, the MENLOPS result
will be as well. Hence, the MENLOPS technique will not impair the accuracy of
the parton shower itself. Higher jet multiplicities exhibit the same accuracy 
as in the \MEPS approach.

\section{Merging at next-to leading order}
\label{Sec:NLOMerging}
The previous section sets the scene to introduce a new prescription, which
consistently merges multiple \MCatNLO-matched event samples of increasing 
jet multiplicity.  The method is constructed such that it is next-to-leading 
order accurate for observables that are sensitive to both $\Phi_n$ and 
$\Phi_{n+1}\,\Theta(Q-Q_\text{cut})$, while maintaining the logarithmic
accuracy of \MCatNLO for observables sensitive to $\Phi_{n+1}$. In other words,
the goal is to describe every jet observable at next-to leading order in
the strong coupling constant, including Sudakov suppression factors.

\subsection{Definition of the \protect\MEPSatNLO technique}
\label{Sec:mepsatnlo}
We propose a method based on the following expression for the 
expectation value of an arbitrary infrared-finite observable $O$
\begin{equation}\label{eq:nlomerging}
  \begin{split}
  \abr{O}\,=&\;
  \int\done\Phi_n\;\bar{\rm B}_n^\text{(A)}
    \Bigg[
	\Delta_n^\text{(A)}(t_c,\mu_Q^2)\,O_n
        +\int\limits_{t_c}^{\mu_Q^2}\done\Phi_1\;
         \frac{{\rm D}_n^\text{(A)}}{{\rm B}_n}\,
	 \Delta_n^\text{(A)}(t_{n+1},\mu_Q^2)\,\Theta(Q_\text{cut}-Q_{n+1})\;
         O_{n+1}\;
  \,\Bigg]\\
  &\hspace*{-5mm}
    +\int\done\Phi_{n+1}\;\mr{H}_n^{\rm(A)}\,
    \Delta_n^\text{(PS)}(t_{n+1},\mu_Q^2)\,
   \Theta(Q_\text{cut}-Q_{n+1})\;O_{n+1}\\
  &\hspace*{-5mm}
   +\int\done\Phi_{n+1}\;\bar{\rm B}_{n+1}^\text{(A)}
   \Bigg(\,
     1+\frac{{\rm B}_{n+1}}{\bar{\rm B}_{n+1}^\text{(A)}}
     \int\limits_{t_{n+1}}^{\mu_Q^2}\done\Phi_1\,{\rm K}_n\,
     \Bigg)
   \Delta_n^\text{(PS)}(t_{n+1},\mu_Q^2)\,\Theta(Q_{n+1}-Q_\text{cut})\\
  &\hspace*{15mm}
    \times\Bigg[
	\Delta_{n+1}^\text{(A)}(t_c,t_{n+1})\,O_{n+1}
        +\int\limits_{t_c}^{t_{n+1}}\done\Phi_1\;
         \frac{{\rm D}_{n+1}^\text{(A)}}{{\rm B}_{n+1}}\,
	 \Delta_{n+1}^\text{(A)}(t_{n+2},t_{n+1})\;
         O_{n+2}\;
  \,\Bigg]\\
  &\hspace*{-5mm}
  +\int\done\Phi_{n+2}\;\mr{H}_{n+1}^{\rm(A)}\,
  \Delta_{n+1}^\text{(PS)}(t_{n+2},t_{n+1})\,\Delta_n^\text{(PS)}(t_{n+1},\mu_Q^2)\,
  \Theta(Q_{n+1}-Q_\text{cut})\;O_{n+2}\;+\,\ldots\;,
  \end{split}
\end{equation}
where again the obvious phase space arguments in the matrix element
contributions and splitting kernels have been suppressed for better 
readability, and where they have been moved to subscripts in the observables.
The dots indicate contributions from higher parton-level multiplicities, which
are dealt with in an iterative procedure as detailed in Sec.~\ref{Sec:iter}.

The square bracket on the first line and third line is generated by weighted 
parton showers, as discussed in Sec.~\ref{Sec:mc_color}, while all Sudakov factors
$\Delta^{\rm(PS)}$ are generated by standard shower algorithms. The terms 
$\done\Phi_n \bar{\mr{B}}_n^{\rm(A)}$ and $\done\Phi_{n+1} \mr{H}_n^{\rm(A)}$ 
correspond to the fixed-order seed events. A convenient Monte-Carlo algorithm 
to generate the factor $\mr{B}_n/\bar{\mr{B}}_n^{\rm(A)}$ will be discussed 
in Sec.~\ref{sec:mc-implementation}.

It is easy to show that next-to-leading order accuracy is maintained for
observables sensitive to $\Phi_{n+1}$ at $Q>Q_{\rm cut}$, where $Q$ is the 
transverse momentum scale of the first emission, i.e.\ of parton $n+1$. 
Expanding the Sudakov form factor $\Delta_n^{\rm(PS)}(t,\mu_Q^2)$ in the
third line to first order and combining it with the square bracket 
on the same line yields correction terms which are at most of 
$\mc{O}(\alpha_s^2)$.

In order to show the logarithmic accuracy of the procedure, 
Eq.~\eqref{eq:nlomerging} is rewritten as follows
\begin{equation}\label{eq:split_mepsnlo1}
  \abr{O}\,=\;\abr{O}^{\rm MC@NLO}\,+\,\abr{O}^{\rm corr}\,,
\end{equation}
with $\abr{O}^{\rm MC@NLO}$ given by \eqref{eq:mcatnlo}. Taking into account 
only $n+1$ parton final states the correction term this time is given by
\footnote{Additional contributions are at most of $\mc{O}(\alpha_s^2L^2)$ and 
thus do not impair the logarithmic or fixed order accuracy we intend to 
prove.}
\begin{equation}\label{eq:nlomerging_mod}
  \begin{split}
  \abr{O}^{\rm corr}\,=\,&\;
  \int\done\Phi_{n+1}\,\Theta(Q_{n+1}-Q_\text{cut})\;
  O_{n+1}\;\Delta_{n+1}^\text{(PS)}(t_c, t_{n+1})\Delta_n^\text{(PS)}(t_{n+1},\mu_Q^2)\,
  \\&\hspace*{-10mm}\times\,
  \Bigg\{
  \bar{\mr{B}}_{n+1}^\text{(A)}\Bigg(1+
  \frac{\mr{B}_{n+1}}{\bar{\mr{B}}_{n+1}^\text{(A)}}
  \int\limits_{t_{n+1}}^{\mu_Q^2}\done\Phi_1\,{\rm K}_n\Bigg)
  \frac{\Delta_{n+1}^{\rm(A)}(t_c,t_{n+1})}{\Delta_{n+1}^{\rm(PS)}(t_c,t_{n+1})}\,-\,
  \mr{H}^\text{(A)}_n\,-\,
  \frac{\bar{\mr{B}}_{n}^\text{(A)}}{\mr{B}_{n}}\,\mr{D}^\text{(A)}_n\,
  \frac{\Delta_n^\text{(A)}(t_{n+1},\mu_Q^2)}{\Delta_n^\text{(PS)}(t_{n+1},\mu_Q^2)}
  \Bigg\}\\
  \,=\,&
  \int\done\Phi_{n+1}\,\Theta(Q_{n+1}-Q_\text{cut})\;
  O_{n+1}\;\Delta_n^\text{(PS)}(t_{n+1},\mu_Q^2)\,
  \\&\hspace*{-10mm}\times\,
  \Bigg\{
  \mr{D}^\text{(A)}_n\,\Bigg[1-
    \frac{\bar{\mr{B}}_{n}^\text{(A)}}{\mr{B}_{n}}\,
    \frac{\Delta_n^\text{(A)}(t_{n+1},\mu_Q^2)}
         {\Delta_n^\text{(PS)}(t_{n+1},\mu_Q^2)}\Bigg]\\
  &\hspace*{-3mm}
  -\mr{B}_{n+1}\Bigg[
    1\,-\,\Bigg(
    \frac{\bar{\mr{B}}_{n+1}^\text{(A)}}{\mr{B}_{n+1}}
    \,+\,\int\limits_{t_{n+1}}^{\mu_Q^2}\done\Phi_1\,{\rm K}_n\Bigg)
    \frac{\Delta_{n+1}^\text{(A)}(t_c,t_{n+1})}
         {\Delta_{n+1}^\text{(PS)}(t_c,t_{n+1})}
    \Bigg]
  \Bigg\}\;.
  \end{split}
\end{equation}

Both terms in the curly brackets consist of one factor describing
the emission of an extra particle,
$\mr{D}^\text{(A)}_{n}$ and $\mr{B}_{n+1}$. Those will eventually yield 
a contribution of $\mc{O}(\alpha_sL^2)$.  The factors multiplying
these emission terms are at most of $\mc{O}(\alpha_sL)$.
However, these logarithms, if present, are due to sub-leading colour
configurations stemming from the difference between $\Delta^\text{(A)}$ and $\Delta^\text{(PS)}$.
The combination of virtual and real contributions in $\bar{\mr{B}}^\text{(A)}_{n}$ does not induce any logarithms spoiling the accuracy of the parton shower.
Thus the correction term does not impair the formal logarithmic
accuracy of the parton shower.  

It is worth noting here that the algorithm detailed in~\cite{Lavesson:2008ah}, 
while aiming at the same formal accuracy, follows a different construction 
paradigm.  Rather than starting from the matrix elements, like the approach 
presented here, and matching the showers to them, its authors start from the 
parton shower and correct its emissions with higher order matrix elements.

\subsection{Iteration for multijet events}
\label{Sec:iter}
Having shown, for the case of the first additional emission, how NLO- and the
logarithmic accuracy of the shower are maintained, we now turn to the question
how this can also be shown for the $k$th additional jet.  The first thing to
be understood is that, in general, the observable $O$ will have
support in different sectors by different jet multiplicities.  In the formalism
outlined here this is reflected by the $\Theta$-functions involving the jet
cut $Q_{\rm cut}$ and the scale $Q$ of the softest emission of a given Born-like
$(n+k)$-jet configuration, in general given by $Q_{n+k}=Q(\Phi_{n+k})$.  For such
a configuration, the respective expression for the $(n+k)$-exclusive jet part 
of the observable, 
\begin{equation}\label{eq:exclusiveO}
\abr{O}_{n+k}^{\rm excl} = 
\sum\limits_{j=n+k}^\infty \abr{O_j\,\Theta(Q_{n+k}-Q_{\rm cut})\Theta(Q_{\rm cut}-Q_{n+k+1})}\,, 
\end{equation}
is given by the suitably modified second part of Eq.\ (\ref{eq:nlomerging}), 
\begin{equation}\label{eq:nlo_term}
  \begin{split}
    \abr{O}_{n+k}^{\rm excl}\,=&\;
    \int\done\Phi_{n+k}\;
    \Theta(Q_{n+k}-Q_{\rm cut})\,\bar{\rm B}_{n+k}^\text{(A)}\,
    \\&\hspace*{-7mm}
    \times\,\Bigg[\,\prod\limits_{i=n}^{n+k-1}\,\Delta_{\,i}^{\rm(PS)}(t_{i+1},t_i)\,
    \Bigg(\,1+\frac{\mr{B}_{n+k}}{\bar{\mr{B}}_{n+k}^\text{(A)}}
    \int\limits_{t_{i+1}}^{t_i}\done\Phi_1\,{\rm K}_i\,
    \Bigg)\Bigg]
    \\&\hspace*{-7mm}
    \times\;\Bigg[\Delta_{n+k}^\text{(A)}(t_c,t_{n+k})\,O_{n+k}\,+\,
      \int\limits_{t_c}^{t_{n+k}}\done\Phi_1\;
      \frac{\mr{D}_{n+k}^\text{(A)}}{\mr{B}_{n+k}}\,
      \Delta_{n+k}^\text{(A)}(t_{n+k+1},t_{n+k})
      \Theta(Q_\text{cut}-Q_{n+k+1})\;O_{n+k+1}\,\Bigg]
    \\&\hspace*{-10mm}
    +\int\done\Phi_{n+k+1}\;\Theta(Q_{n+k}-Q_{\rm cut})\,
    \Theta(Q_\text{cut}-Q_{n+k+1})\;O_{n+k+1}
    \mr{H}_{n+k}^{\rm(A)}\,
    \prod\limits_{i=n}^{n+k}\,\Delta_{\,i}^{\rm(PS)}(t_{i+1},t_i)\,.
  \end{split}
\end{equation}

In order to see the formal accuracy of this expression, let us define an 
$(n+k)$-jet inclusive expression of the observable, by dropping the second 
$\Theta$-function in \eqref{eq:exclusiveO}.  As before, it can be written as 
the sum of an \MCatNLO-like expression acting on the $(n+k)$-parton Born 
configuration and a correction term,
\begin{equation}
  \abr{O}_{n+k}^{\rm incl}\;=\;
  \abr{O}_{n+k}^{\rm MC@NLO}\,+\,\abr{O}_{n+k}^{\rm corr}\,,
\end{equation}
where 
\begin{equation}
  \begin{split}
    \abr{O}_{n+k}^{\rm MC@NLO}\,=&\;
    \int\done\Phi_{n+k}\;
    \Theta(Q_{n+k}-Q_{\rm cut})\,\bar{\rm B}_{n+k}^\text{(A)}\,
    \\&\hspace*{-12mm}
    \times\,\Bigg[\,\prod\limits_{i=n}^{n+k-1}\,\Delta_{\,i}^{\rm(PS)}(t_{i+1},t_i)\,
    \Bigg(\,1+\frac{{\rm B}_{n+k}}{\bar{\rm B}_{n+k}^\text{(A)}}
    \int\limits_{t_{i+1}}^{t_i}\done\Phi_1\,{\rm K}_i\,
    \Bigg)\Bigg]\\
    &\hspace*{-12mm}
    \times\,
    \Bigg[\Delta_{n+k}^\text{(A)}(t_c,t_{n+k})\,O_{n+k}
      +\int\limits_{t_c}^{t_{n+k}}\done\Phi_1\;
      \frac{{\rm D}_{n+k}^\text{(A)}}{{\rm B}_{n+k}}\,
      \Delta_{n+k}^\text{(A)}(t_{n+k+1},t_{n+k})\,
      O_{n+k+1}\;\Bigg]\\
    &\hspace*{-15mm}
    +\,\int\done\Phi_{n+k+1}\;
    \Theta(Q_{n+k}-Q_{\rm cut})\;O_{n+k+1}\,
    \mr{H}_{n+k}^{\rm(A)}\,
      \prod\limits_{i=n}^{n+k}\,\Delta_{\,i}^{\rm(PS)}(t_{i+1},t_i)\,.
  \end{split}
\end{equation}
The only difference with respect to the usual form of the \MCatNLO expression 
in \eqref{eq:mcatnlo} is the term in the second line which encodes a veto 
on emissions into the jet region from intermediate lines with its 
$\mc{O}(\alpha_s)$-part subtracted.

At the relevant order in $\alpha_s$, this correction term reads
\begin{equation}\label{eq:proof_int}
  \begin{split}
  \abr{O}_{n+k}^{\rm corr}\,=& \;\int\done\Phi_{n+k+1}\,
  \Theta(Q_{n+k+1}-Q_\text{cut})\;O_{n+k+1}\,
  \prod\limits_{i=n}^{n+k+1}\,\Delta_{\,i}^{\rm(PS)}(t_{i+1},t_i)\,
  \\&\hspace*{-10mm}\times
  \Bigg\{
    \mr{D}_{n+k}^\text{(A)}\Theta(t_{n+k}-t_{n+k+1})\;\Bigg[\,1\,-\,\Bigg(
    \frac{\bar{\mr{B}}_{n+k}^\text{(A)}}{\mr{B}_{n+k}}
    +\sum\limits_{i=n}^{n+k-1}\int\limits_{t_{i+1}}^{t_i}\done\Phi_1\mr{K}_i\Bigg)
    \frac{\Delta_{n+k}^\text{(A)}(t_{n+k+1},t_{n+k})}
         {\Delta_{n+k}^\text{(PS)}(t_{n+k+1},t_{n+k})}\,\Bigg]
  \\&\hspace*{-5mm}
    -\mr{B}_{n+k+1}
    \Bigg[\,1\,-\,
      \Bigg(\frac{\bar{\mr{B}}_{n+k+1}^\text{(A)}}{\mr{B}_{n+k+1}}\,+
    \,\sum\limits_{i=n}^{n+k}\;\int\limits_{t_{i+1}}^{t_i}\done\Phi_1\mr{K}_i\Bigg)
      \frac{\Delta_{n+k+1}^\text{(A)}(t_c,t_{n+k+1})}{
          \Delta_{n+k+1}^\text{(PS)}(t_c,t_{n+k+1})}\,\Bigg]\,
    \Bigg\}\;,
  \end{split}
\end{equation}
and the same reasoning already applied to Eq.~\eqref{eq:nlomerging_mod} yields
the desired result.  For a more detailed discussion, including the effect
of truncated showering, see \cite{Hoeche:2012yf}.  

The finding above shows that no terms appear due to the merging prescription 
that violate the logarithmic accuracy of the parton shower at and around 
$Q_{\rm cut}$.  To see this, it is sufficient to analyse the first emission off 
the $(n+k)$-jet configuration over the full phase space.  The second emission 
is, of course, completely determined by the parton shower and thus correct 
by definition.  Also, clearly, the phase space for this first emission is 
confined to the region below $Q_{\rm cut}$, therefore the behaviour above
this scale is defined by the parton-level result with next higher multiplicity, 
the $(n+k+1)$-jet configuration.  By however extending the first emission above
this cut and analysing the impact on $\mc{O}_{n+k+1}$ we show that the two
regions match as smoothly as the logarithmic accuracy of the parton shower
dictates.

\subsection{Renormalisation scale uncertainties}
\label{Sec:renorm}
The key aim of the \MEPSatNLO approach presented here is to reduce the dependence of the
merged prediction on the renormalisation scale $\mu_R$, which is employed in the computation
of the hard matrix elements. This scale has not been made explicit so far.

Note that only the dependence on the renormalisation scale is reduced compared to the
\MEPS method, while the dependence on the resummation scale, $\mu_Q$, remains the same.
This is a direct consequence of the fact that the parton-shower evolution is not improved
in our prescription, but only the accuracy of the hard matrix elements.
The resummation scale dependence was analysed in great detail in~\cite{Hoeche:2011fd}. 

Following the \MEPS strategy, the renormalisation scale should be determined by analogy 
of the leading-order matrix element with the respective parton shower branching 
history~\cite{Hoeche:2009rj}. In next-to-leading order calculations, however, one needs 
a definition which is independent of the parton multiplicity. The same scale should be used 
in Born matrix elements and real-emission matrix elements if they have similar kinematics, 
and in particular when the additional parton of the real-emission correction becomes soft or collinear. 
This can be achieved if we define the renormalisation scale for a process of $\mc{O}(\alpha_s^n)$ as \cite{Hamilton:2012np}
\begin{equation}\label{eq:mur_def}
  \alpha_s(\mu_R^2)^n\,=\;\prod_{i=1}^n\alpha_s(\mu_i^2)\;,
\end{equation}
a procedure that has been used in LO merging for some time.  
Here, $\mu_i^2$ are the respective scales defined by analogy of the Born 
configuration with a parton-shower branching history\footnote{
  In the case of the real-emission correction and the corresponding dipole 
  subtraction terms we consider the underlying Born configuration instead.
  The same scale definition is used in the parton shower and, consequently, 
  in the Sudakov form factors.  Of course, the nodal scales $\mu_i$ found in 
  the backward clustering on the Born-like configuration of a single event 
  then enter the truncated showering.}.  

The renormalisation scale uncertainty in the \MEPSatNLO approach is then determined
by varying $\mu_R\to\tilde{\mu}_R$, while simultaneously correcting for the one-loop effects induced by a
redefinition in Eq.~\eqref{eq:mur_def}. That is, the Born matrix element is multiplied by
\begin{equation}\label{eq:mur_redef}
  \alpha_s(\tilde{\mu}_R^2)^n\,\Bigg(1-\frac{\alpha_s(\tilde{\mu}_R^2)}{2\pi}\,
    \beta_0\sum_{i=1}^n\log\frac{\mu_i^2}{\tilde{\mu}_R^2}\Bigg)\;,
\end{equation}
to generate the one-loop counter-term, while higher-order contributions remain the same.

\section{Monte-Carlo implementation}
\label{sec:mc-implementation}

In this section we describe the Monte Carlo implementation of the merging formula
Eq.~\eqref{eq:nlomerging} in \Sherpa. The techniques needed to combine leading-order 
matrix elements with parton showers are given elsewhere~\cite{Hoeche:2009rj}.

\subsection{Generation of the parton-shower counterterm}
In addition, we now have to implement a method to generate the parton-shower counterterm 
on the third line of Eq.~\eqref{eq:nlomerging}. Note that, by construction, 
this counterterm has the same functional form as the exponent 
of the Sudakov form factor $\Delta^{\rm(PS)}_n(t,\mu_Q^2)$. We can therefore use 
the following algorithm:
\begin{itemize}
\item Start from an $n$-parton configuration underlying the $n+1$-parton event
  at scale $\mu_Q^2$,\\ and implement a truncated parton shower with lower cutoff scale $t$.
\item If no emission is produced, the original $n+1$-parton configuration is retained.
\item If the first emission is generated at scale $t'$ with $Q>Q_{\rm cut}$, the event weight 
  is multiplied\\ by $1/\kappa$, where $\kappa=\bar{\rm B}^{\rm(A)}_{n+1}(\Phi_{n+1})/\mr{B}_{n+1}(\Phi_{n+1})$.
  Evolution is restarted at $t'$.
\item All subsequent emissions are treated as in a standard truncated vetoed parton shower.
\end{itemize}
Events will then be distributed as
\begin{equation}
  \begin{split}
  &\Delta_n^{\rm(PS)}(t,\mu_Q^2)+
  \frac{1}{\kappa}\int_t^{\mu_Q^2}\done\Phi_1\,\Big[\mr{K}_n(\Phi_1)\,\Theta(Q-Q_{\rm cut})\,
    \Delta_n^{\rm(PS)}(t',\mu_Q^2)\,\Big]\Delta_n^{\rm(PS)}(t,t')
  \\&\qquad=\Delta_n^{\rm(PS)}(t,\mu_Q^2)\,
    \Bigg[\,1+\frac{1}{\kappa}\int_t^{\mu_Q^2}\done\Phi_1\,\mr{K}_n(\Phi_1)\,\Theta(Q-Q_{\rm cut})\,\Bigg]\;.
  \end{split}
\end{equation}

This simple algorithm allows to identify the $\mc{O}(\alpha_s)$ counterterm with 
an omitted emission and to generate the correction term on-the-flight, much like
the Sudakov form factor is computed in any parton-shower algorithm itself.

\subsection{Generation of the \protect\MCatNLO Sudakov form factor}
\label{Sec:mc_color}
In this subsection we briefly recall an algorithm to compute \MCatNLO Sudakov 
form factors~\cite{Hoeche:2011fd}, which is one of the basic ingredients to 
our method.

It is well known how to generate emissions according to Sudakov form factors
with strictly negative exponent. In our implementation of \MCatNLO, however,
we have to deal with potentially positive exponents, related to subleading 
colour configurations. This leads to form factors larger than one, which cannot be 
interpreted in terms of no-branching probabilities and which are dealt with using 
a modified Sudakov veto algorithm~\cite{Hoeche:2009xc,Hoeche:2011fd}.

Assume that $f(t)$ is the sole splitting kernel in our parton shower, integrated 
over $z$ and $\phi$. The differential probability for generating a branching 
at scale $t$, when starting from an upper evolution scale $t'$ is then given by
\begin{equation}\label{eq:va_prob}
  \mc{P}(t,t')\,=\;f(t)\,\exp\cbr{-\int_t^{t'}\done\bar{t}\,f(\bar{t})}\;.
\end{equation}
The key point of the veto algorithm is, that even if the primitive of $f(t)$ 
is unknown, one can still generate events according to $\mc{P}$ using an overestimate 
$g(t)\ge f(t)$, if $g(t)$ has a known integral. Firstly, a value $t$ is generated as 
$t=G^{-1}\sbr{\,G(t')+\log \#\,}$. Secondly, the value is accepted with probability 
$f(t)/g(t)$~\cite{Sjostrand:2006za}.

One can now introduce an additional estimate $h(t)$, which is not necessarily an
overestimate of $f(t)$. The related weights are
applied analytically rather than using a hit-or-miss method. They can thus be used
to absorb the negative sign of the \MCatNLO kernels $\mr{D}_n^{\rm(A)}/\mr{B}_n$. 
This leads to a correction factor for one accepted branching with $m$ intermediate 
rejections of
\begin{equation}\label{eq:wva_analytic}
  w(t,t_1,\ldots,t_m)\,=\;\frac{g(t)}{h(t)}\,
    \prod_{i=1}^m\frac{g(t_i)}{h(t_i)}\frac{h(t_i)-f(t_i)}{g(t_i)-f(t_i)}\;,
\end{equation}
where the $t_i$ run over intermediately rejected steps.
Note that Eq.~\eqref{eq:wva_analytic} can lead to negative weights, which reflect the fact
that sub-leading colour configurations are taken into account and that the a-priori
density $h(t)$ might underestimate $f(t)$.

In order to implement an evolution using the \MCatNLO kernels $\mr{D}_n^{\rm(A)}/\mr{B}_n$
we need to identify the function $f$ above with the $(z,\phi)$-integral of these kernels. 
A convenient choice of the function $h$ will be the $(z,\phi)$-integral of the parton-shower 
evolution kernels $\mr{K}_n$. We are then free to choose the auxiliary function $g$ on a 
point-by-point basis, but a convenient way is to define $g=C\,f$, where $C$ is a constant 
larger than one. This guarantees that both acceptance and rejection terms are generated 
in sufficient abundance to reduce statistical fluctuations.

The above method guarantees that all subleading colour single logarithmic corrections
to $\mr{B}_n$ are exponentiated. One can therefore guarantee a process-independent 
exponentiation of next-to-leading colour real-emission corrections in the \MCatNLO.

\section{Results}
\label{Sec:Results}
In this section results obtained with the \MEPSatNLO method are presented 
for the case of $e^+e^-$-annihilation into hadrons. The general-purpose
event generator \Sherpa sets the framework for this study~\cite{Gleisberg:2003xi,Gleisberg:2008ta}.
Leading-order matrix elements are generated with \Amegic~\cite{Krauss:2001iv} and 
\Comix~\cite{Gleisberg:2008fv}. Automated dipole 
subtraction~\cite{Gleisberg:2007md} and the 
Binoth--\-Les Houches interface~\cite{Binoth:2010xt} are employed to 
obtain parton-level events at next-to-leading order with virtual 
corrections provided by the \Blackhat library~\cite{Berger:2008sj,
  Berger:2009ep,Berger:2010vm,Berger:2010zx}.  
The parton shower in \Sherpa is based on Catani-Seymour 
dipole factorisation~\cite{Schumann:2007mg}; the related 
\MCatNLO generator has been presented in~\cite{Hoeche:2011fd}.  In contrast 
to all other \MCatNLO implementations, no leading colour approximation is made
in the first step of the parton shower, cf.\ Sec.~\ref{Sec:mc_color}.  
The resummation scale is determined on an event-by-event basis by backward
clustering as described in~\cite{Hoeche:2009rj}. In the special case of $e^+e^-$
collisions discussed here this simplifies to the centre-of-mass energy.
The results presented here are at the hadron level. Note that the hadronisation 
model in \Sherpa~\cite{Winter:2003tt} has been tuned in conjunction with 
the parton shower and leading order matrix elements. It is therefore not 
surprising when deviations are found in observables that are sensitive to
soft particle dynamics.  In the future this will necessitate a new tune
of the hadronisation based on the NLO-merging outlined here, rather than
on the LO MEPS prescription that has been used so far in \Sherpa.

For each of the inclusive samples discussed in the following we generated
$40\cdot 10^{6}$ weighted events. The sub-contributions in different jet
multiplicities were automatically chosen according to their cross sections.
Within each jet multiplicity, the number of $\mb{H}$-events was statistically
enhanced by a factor of $10$ with respect to the $\mb{S}$-events. The cross
section fraction of negative events was
$1.3\%$ for \MCatNLO, 
$0.4\%$ for \MENLOPS, and 
$10.4\%$ for \MEPSatNLO. 
The generation of $40\cdot 10^{6}$ events needed 1.6 CPU days (\MCatNLO),
1.7 CPU days (\MENLOPS) and 2.0 CPU days (\MEPSatNLO) on Intel Xeon E5440
CPUs at 2.83GHz.

\subsection{Choice of the merging scale}
\begin{figure}
  \begin{center}
    \includegraphics[width=0.5\textwidth]{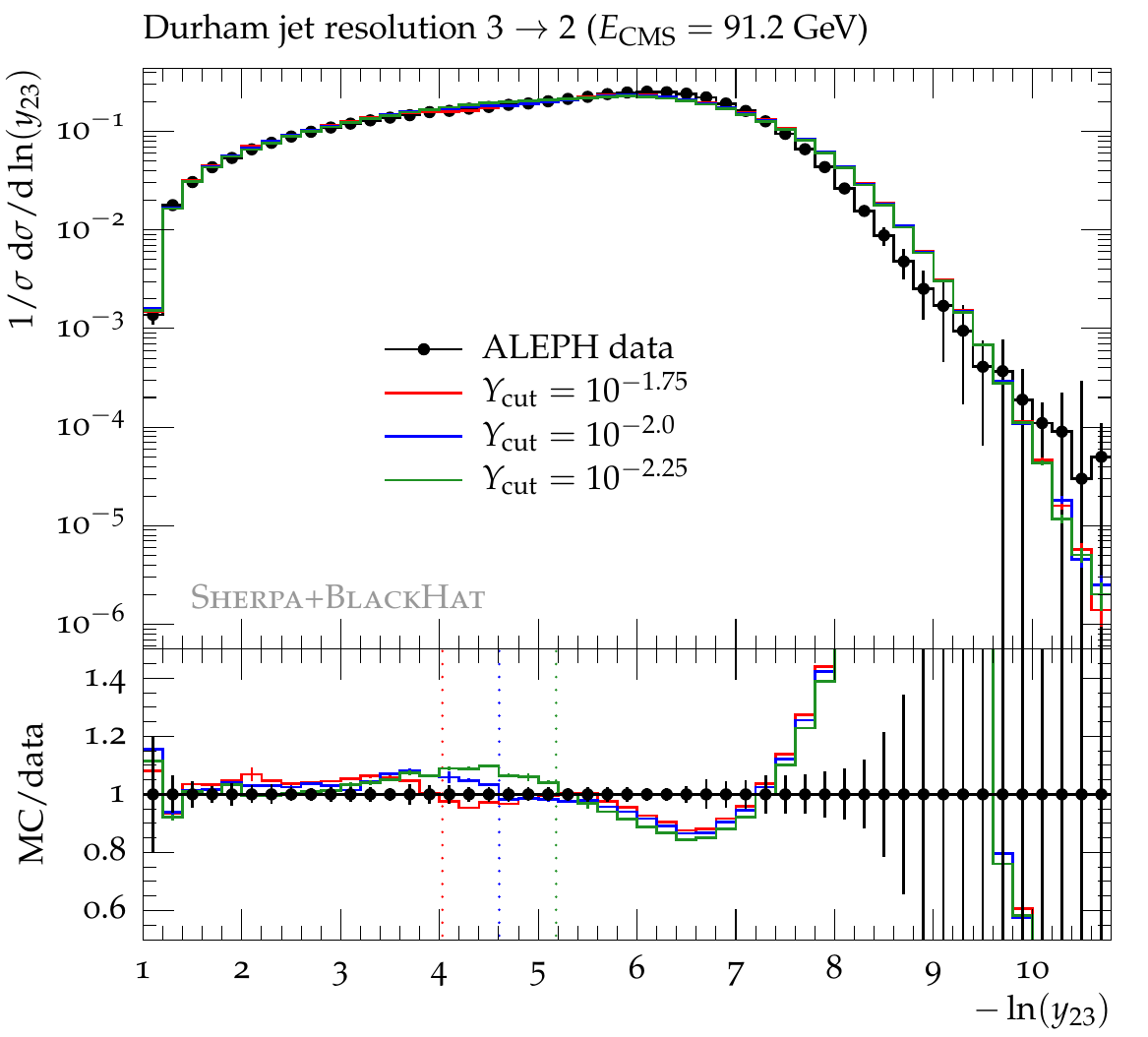}\nolinebreak
    \includegraphics[width=0.5\textwidth]{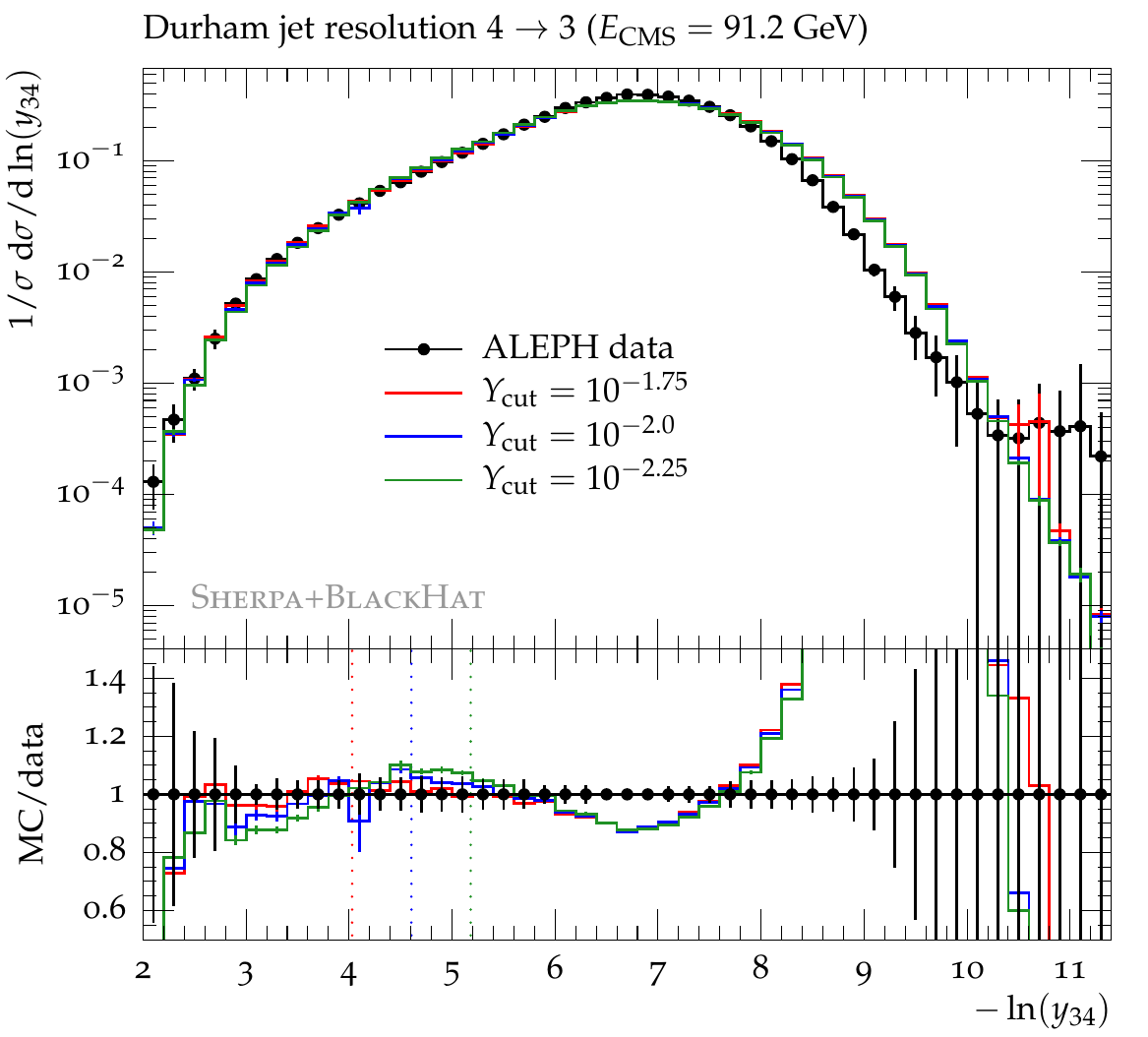}\\
    \includegraphics[width=0.5\textwidth]{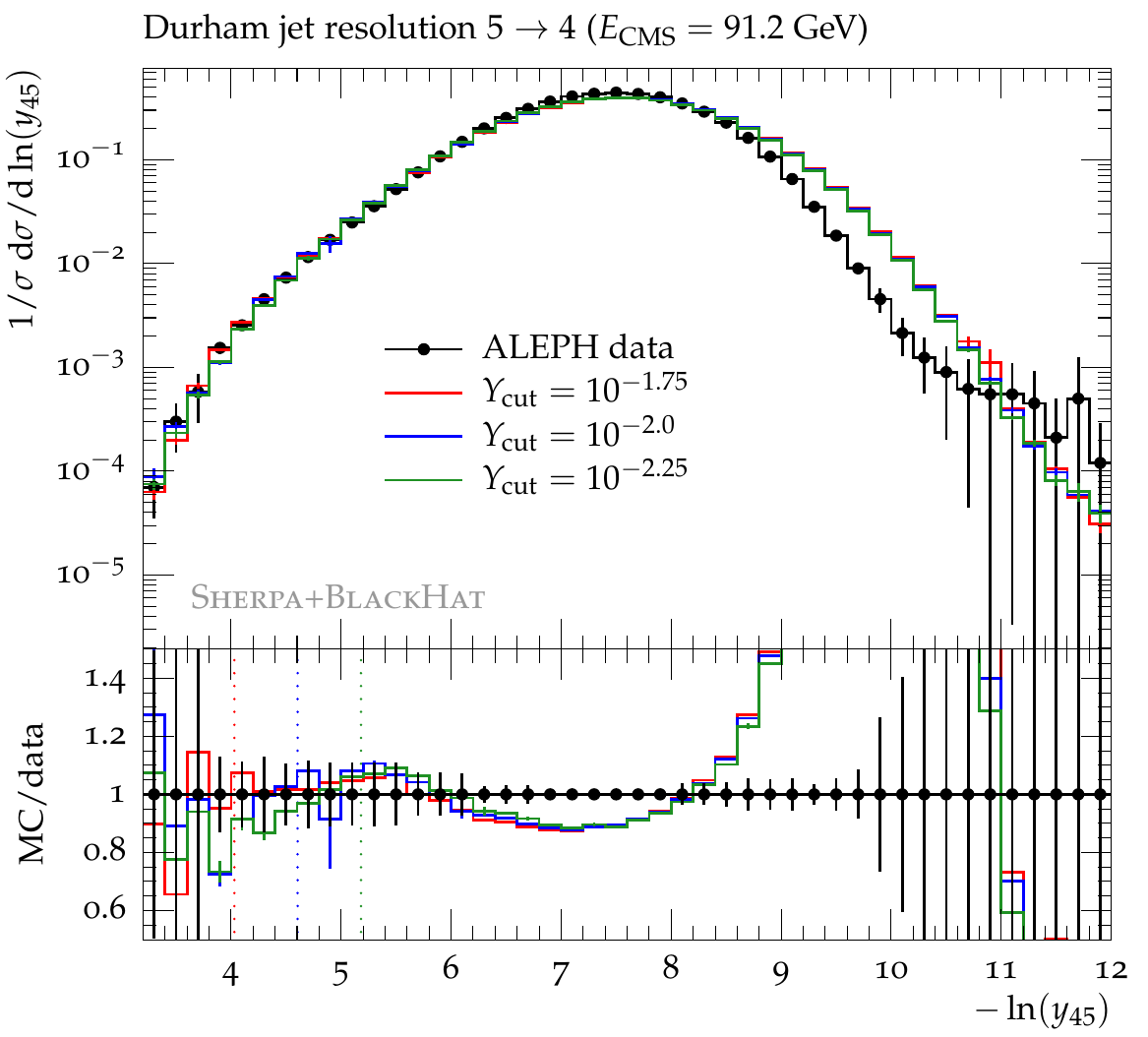}\nolinebreak
    \includegraphics[width=0.5\textwidth]{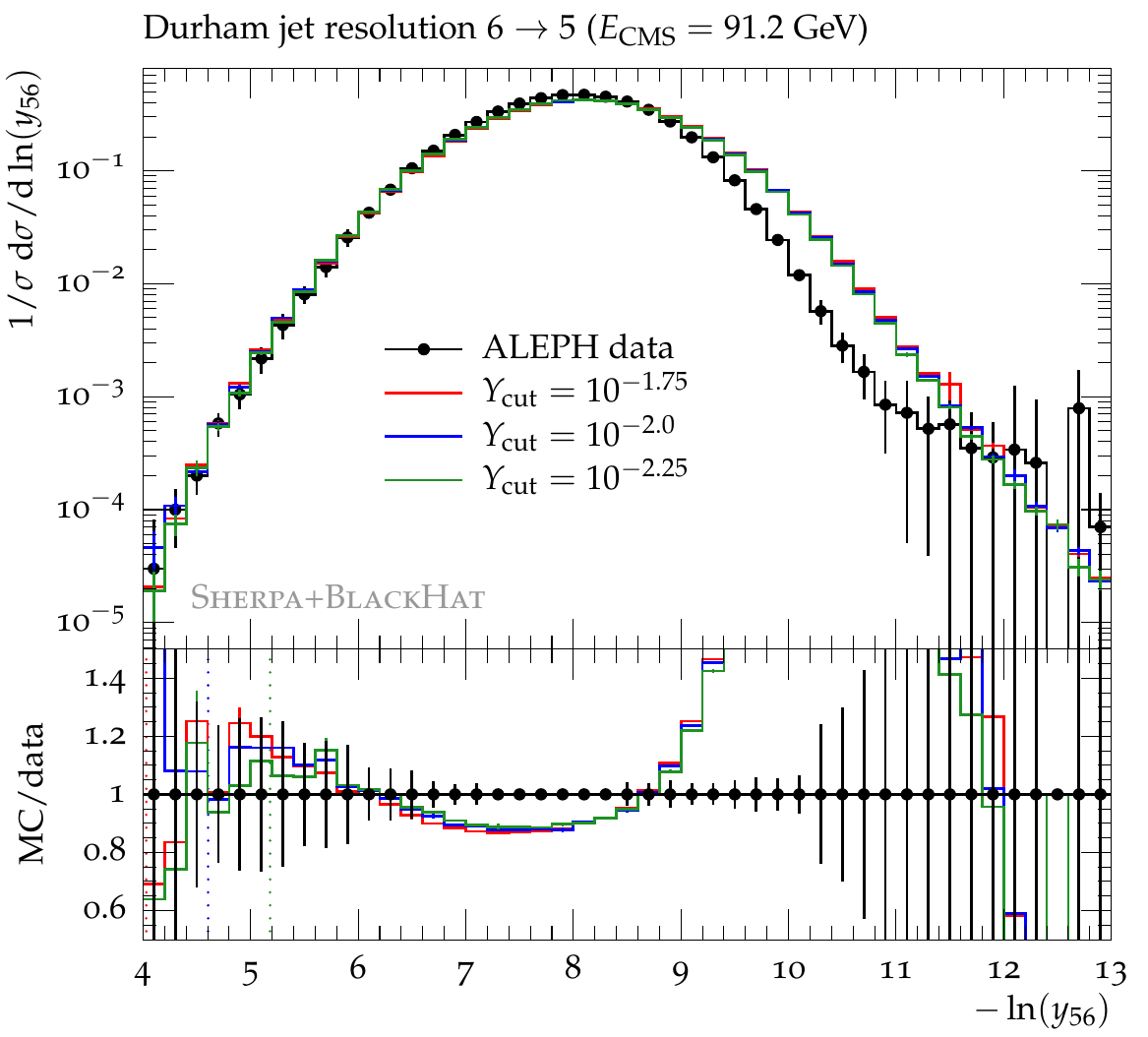}
  \end{center}
  \caption{Experimental data from ALEPH~\cite{Heister:2003aj}
    for the differential $(n+1)\to n$ jet rates with
    $n = \{2,\,3,\,4,\,5\}$ (upper and lower panel, left to right) at the 
    $Z$ pole ($E_{\rm c.m.}=91.2$ GeV) are compared with \protect\MEPSatNLO 
    simulations with different values of the merging cut, 
    $Y_{\rm cut} = 10^{-\{1.75,\,2.0,\,2.25\}}$.  To guide the eye, the merging cuts 
    have been indicated with dotted lines in the same colour in the ratio
    plot.}
  \label{fig:qcutvar}
\end{figure}

Figure~\ref{fig:qcutvar} shows the dependence of \MEPSatNLO predictions for 
the Durham jet resolution on the merging scale $Q_{\rm cut}$. In order to match 
the customary notation we quote $Y_{\rm cut}=(Q_{\rm cut}/E_{\rm cms})^2$.
All results were generated using 2-,3- and 4-jet NLO parton-level calculations
combined with 5- and 6-jet at leading order.  The 
variation of results with $Y_{\rm cut}$ in the region below and around 
$Y_{\rm cut}$ is of the order of 10\%, the predictions above the cut 
are remarkably stable and match the experimental data very well.
Consequently, one should always choose the merging cut such that the analysis 
region is fully contained in the region covered by the NLO calculation of 
interest.

\subsection{Comparison of approaches and their perturbative uncertainties}
\begin{figure}
  \begin{center}
    \includegraphics[width=0.42\textwidth]{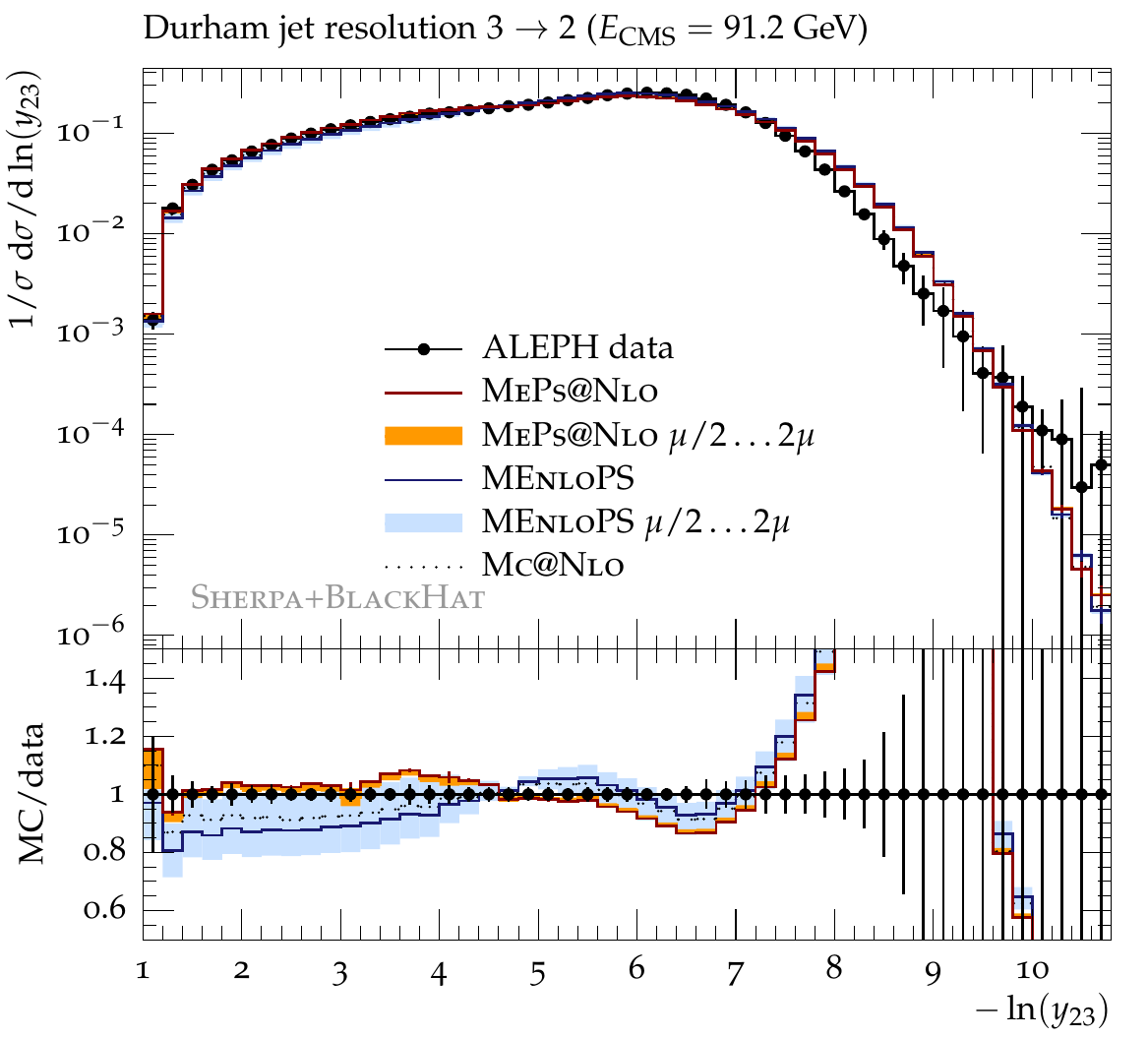}\nolinebreak
    \includegraphics[width=0.42\textwidth]{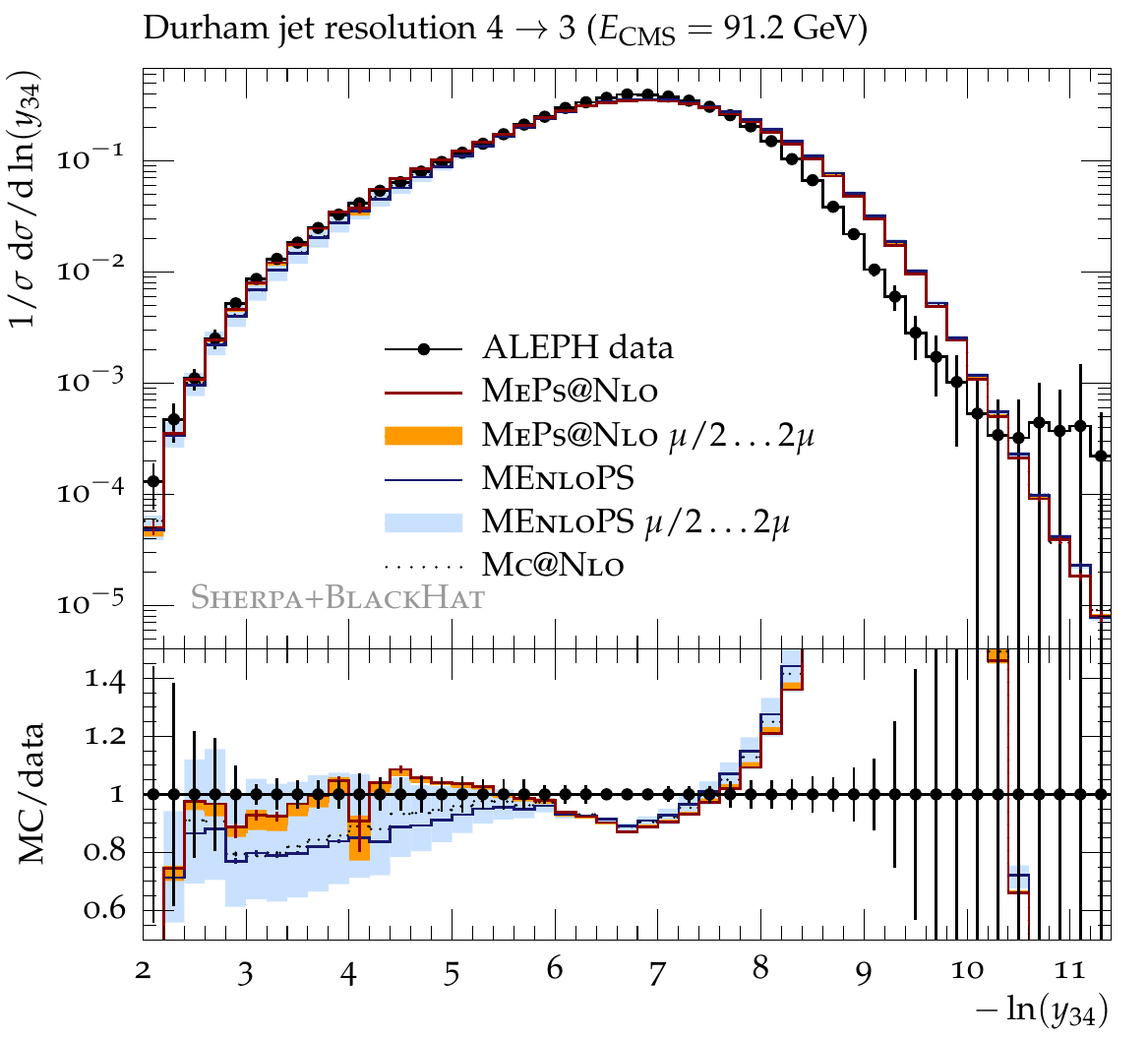}\\
    \includegraphics[width=0.42\textwidth]{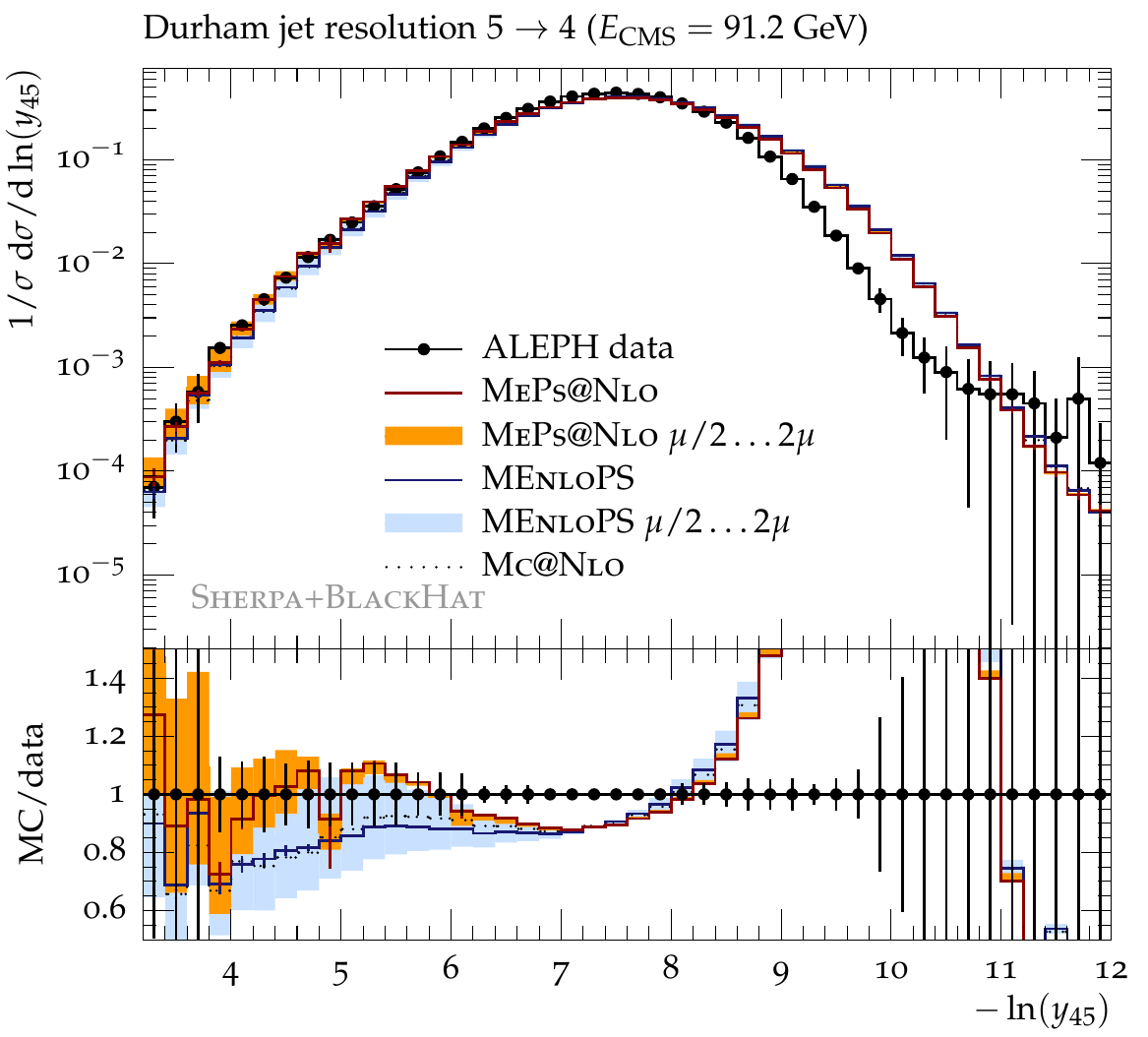}\nolinebreak
    \includegraphics[width=0.42\textwidth]{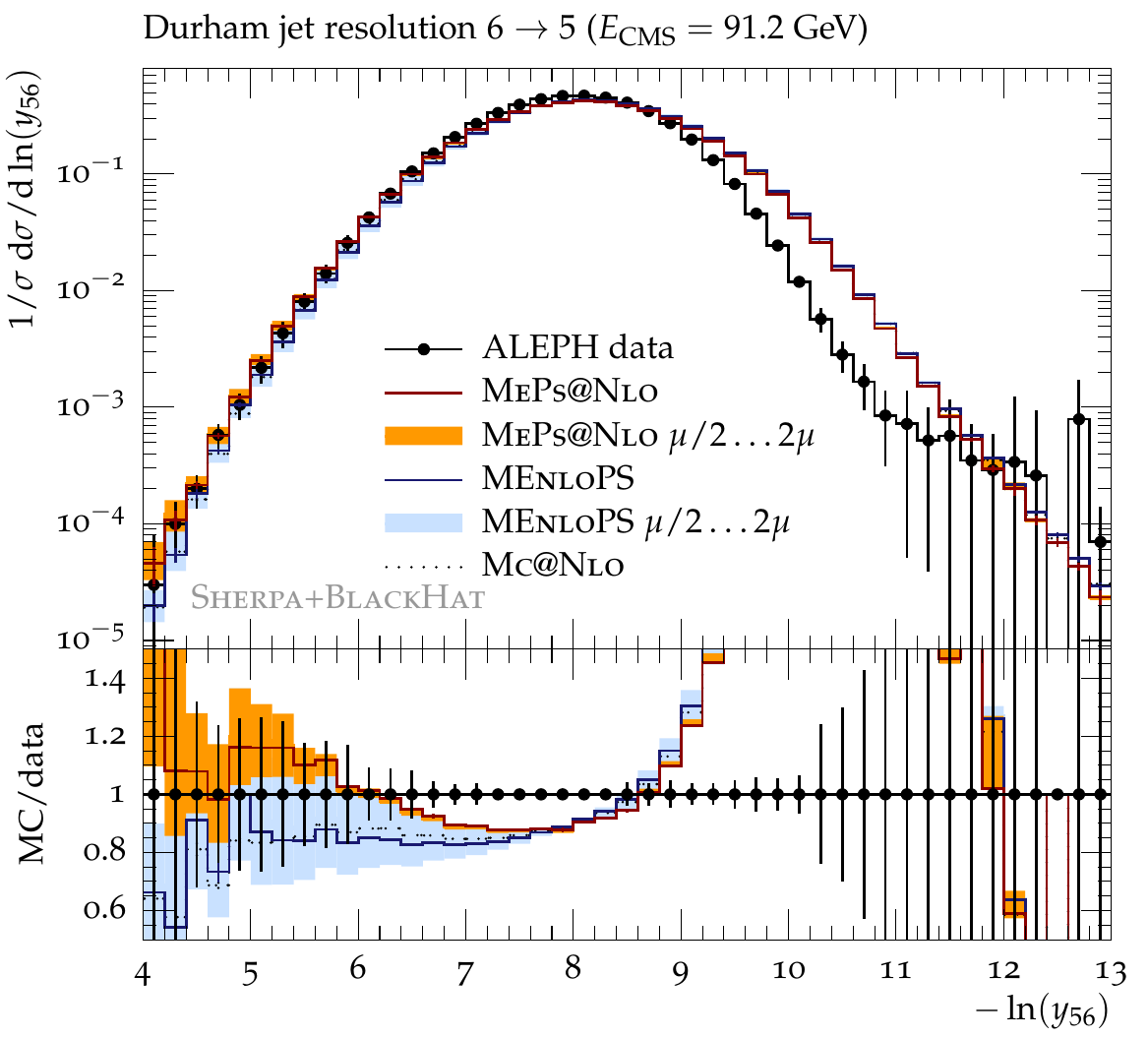}
  \end{center}
  \caption{Perturbative uncertainties in \protect\MENLOPS and \protect\MEPSatNLO 
           predictions of differential jet rates compared to data from 
           ALEPH~\cite{Heister:2003aj}.}
  \label{fig:jetrates}
\end{figure}

\begin{figure}
  \begin{center}
    \includegraphics[width=0.42\textwidth]{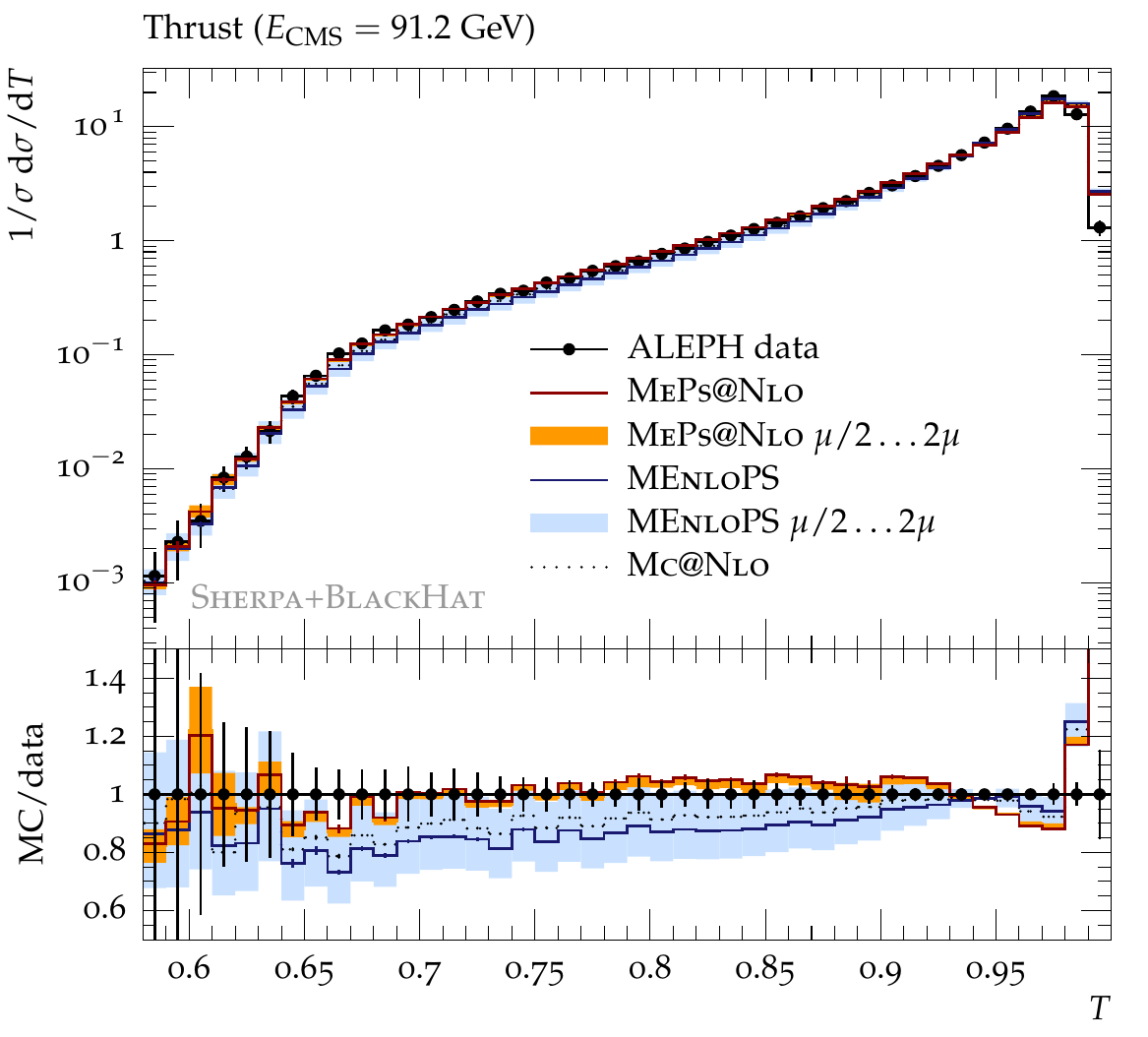}\nolinebreak
    \includegraphics[width=0.42\textwidth]{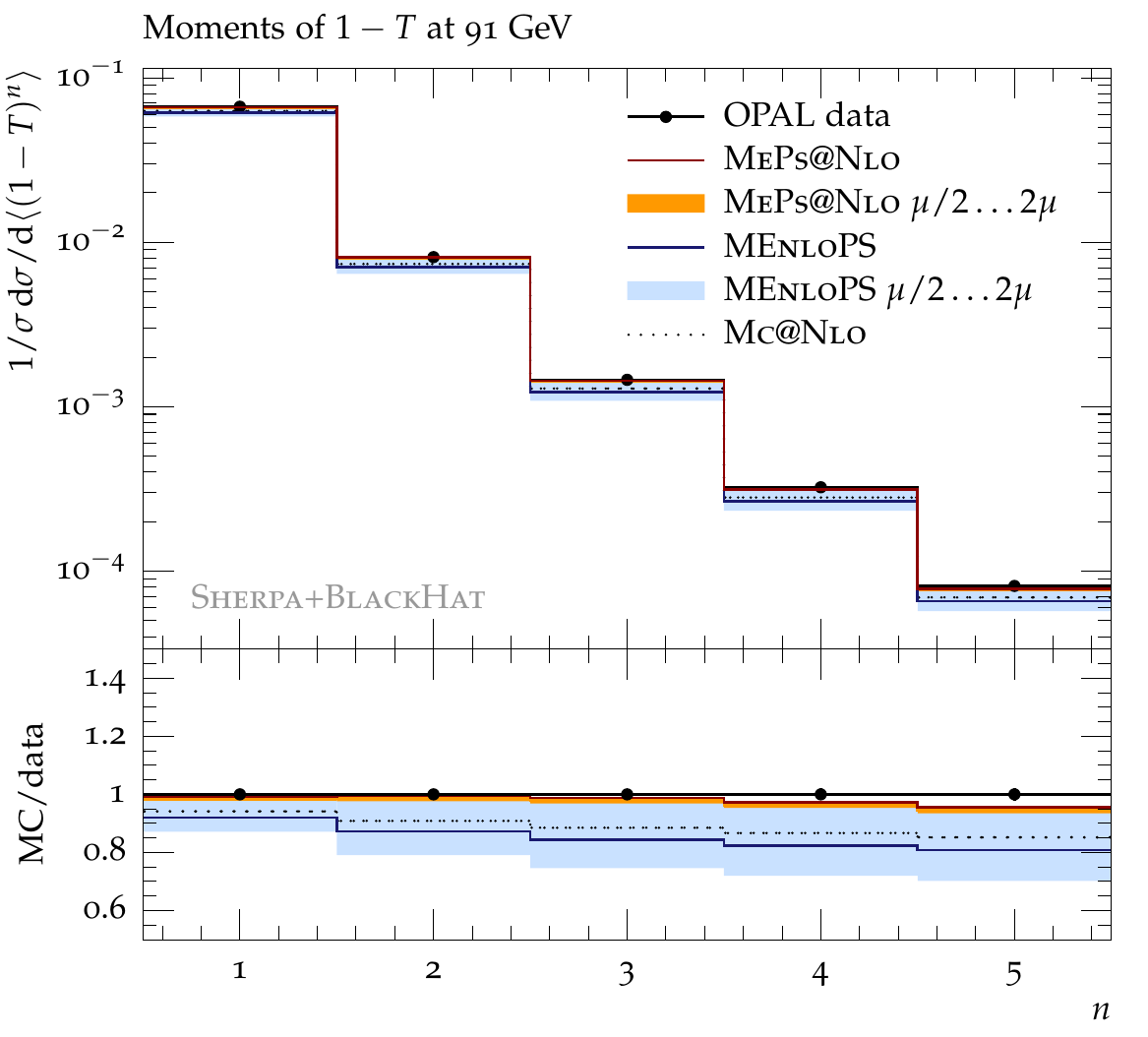}
  \end{center}
  \caption{Perturbative uncertainties in \protect\MENLOPS and \protect\MEPSatNLO 
           predictions of thrust. Compared are the measurements for the event 
           shape from ALEPH~\cite{Heister:2003aj} and its moments from 
           OPAL~\cite{Abbiendi:2004qz}.}
  \label{fig:thrust}
\end{figure}

\begin{figure}
  \begin{center}
    \includegraphics[width=0.42\textwidth]{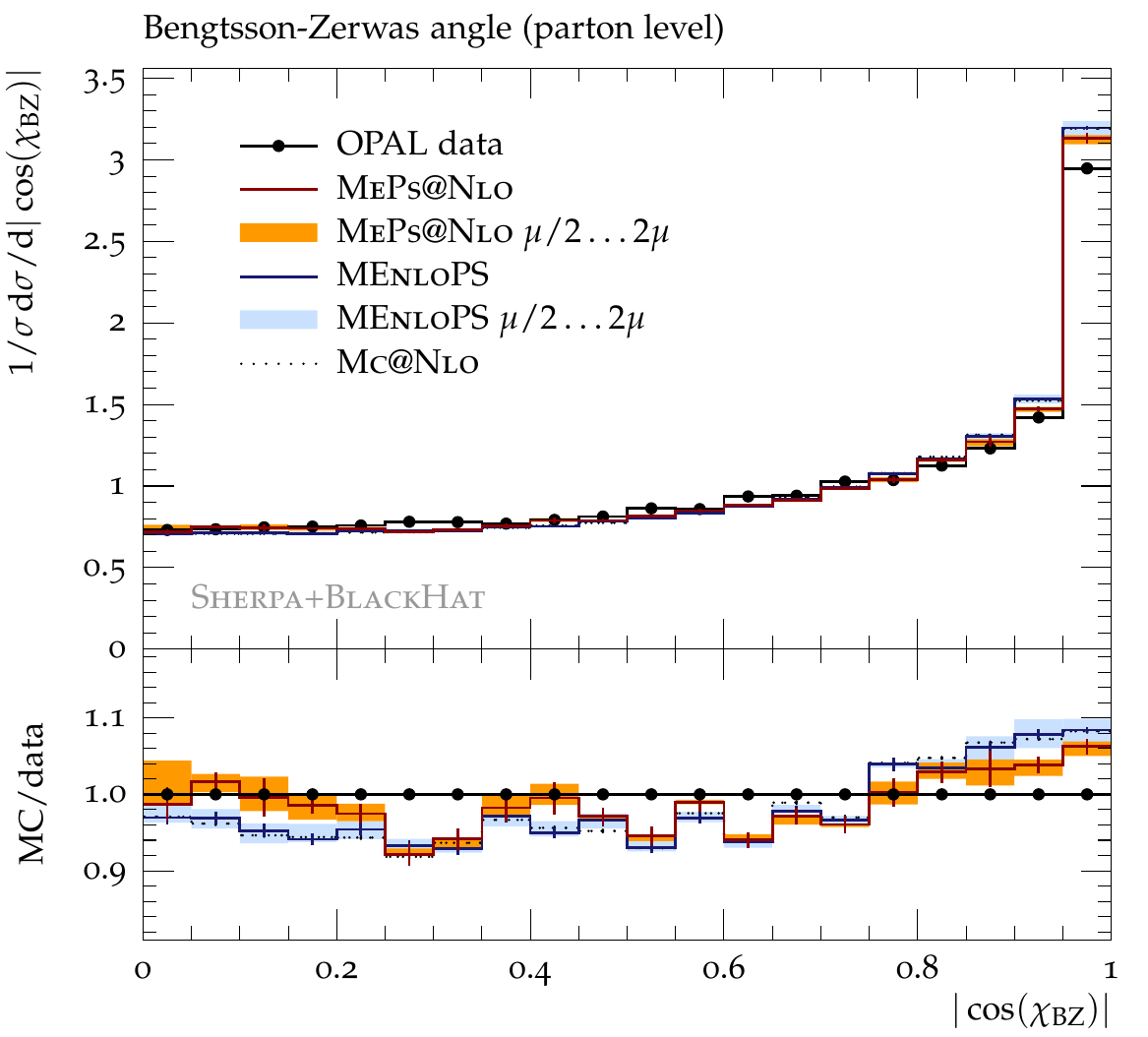}\nolinebreak
    \includegraphics[width=0.42\textwidth]{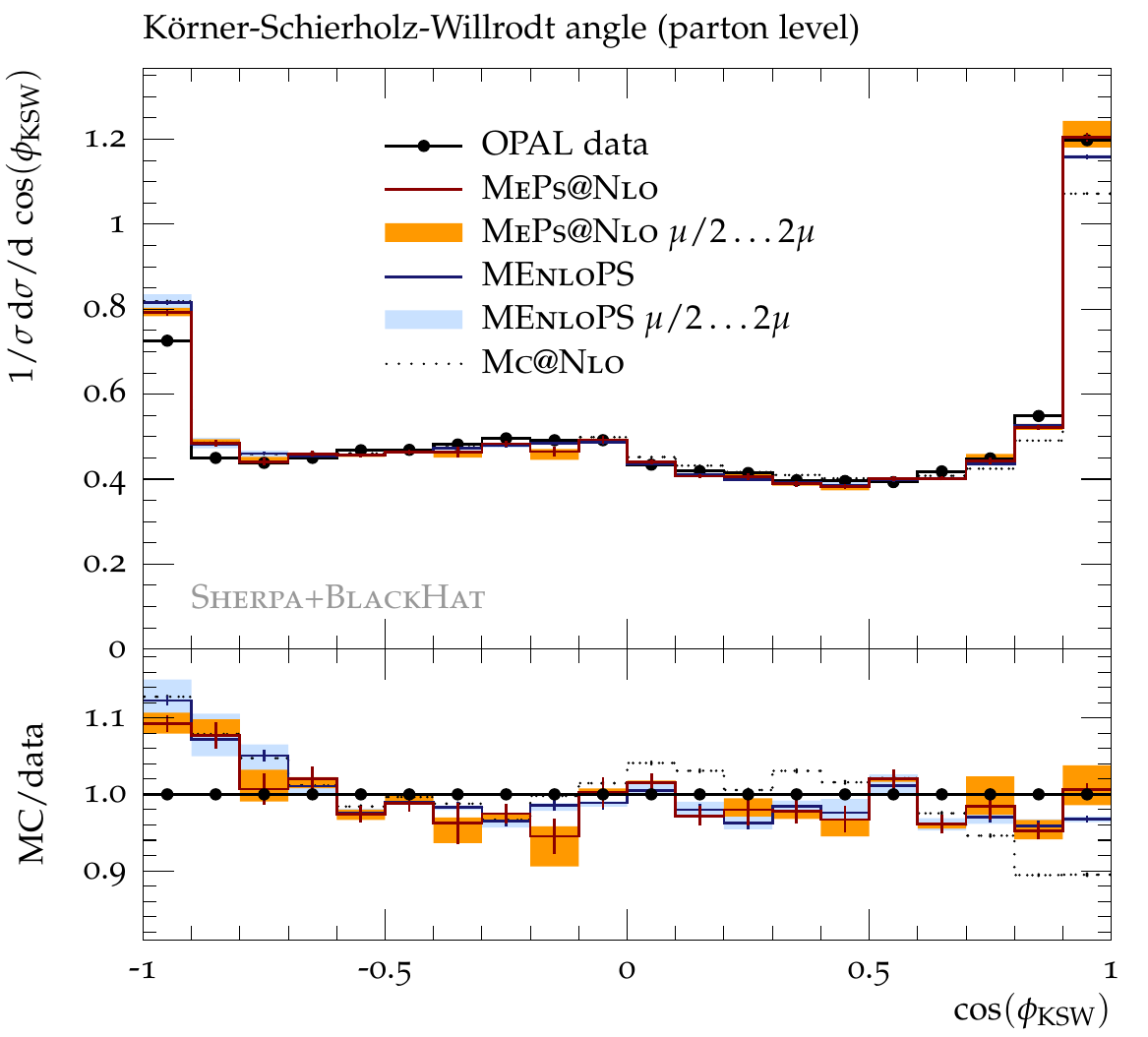}\\
    \includegraphics[width=0.42\textwidth]{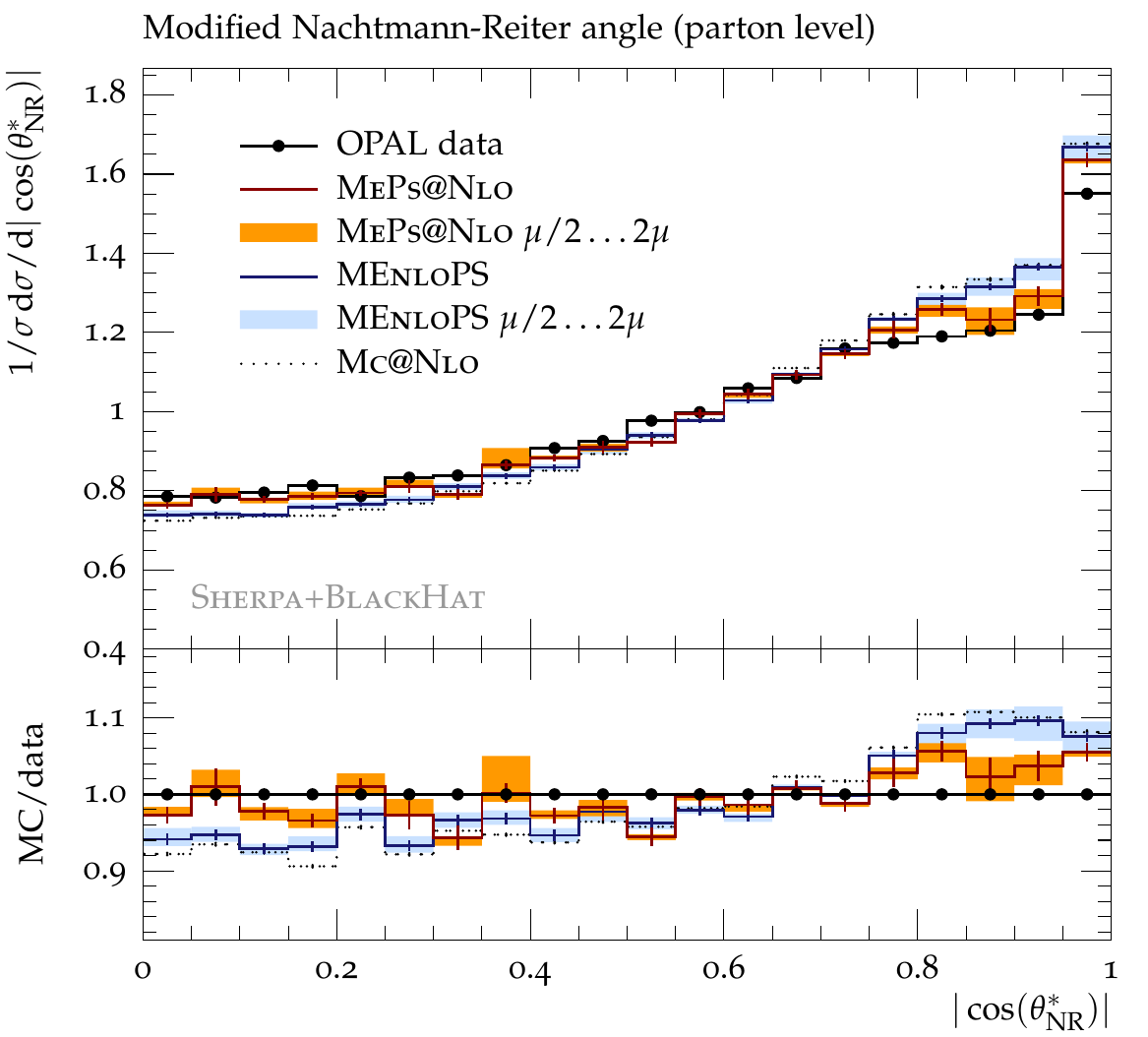}\nolinebreak
    \includegraphics[width=0.42\textwidth]{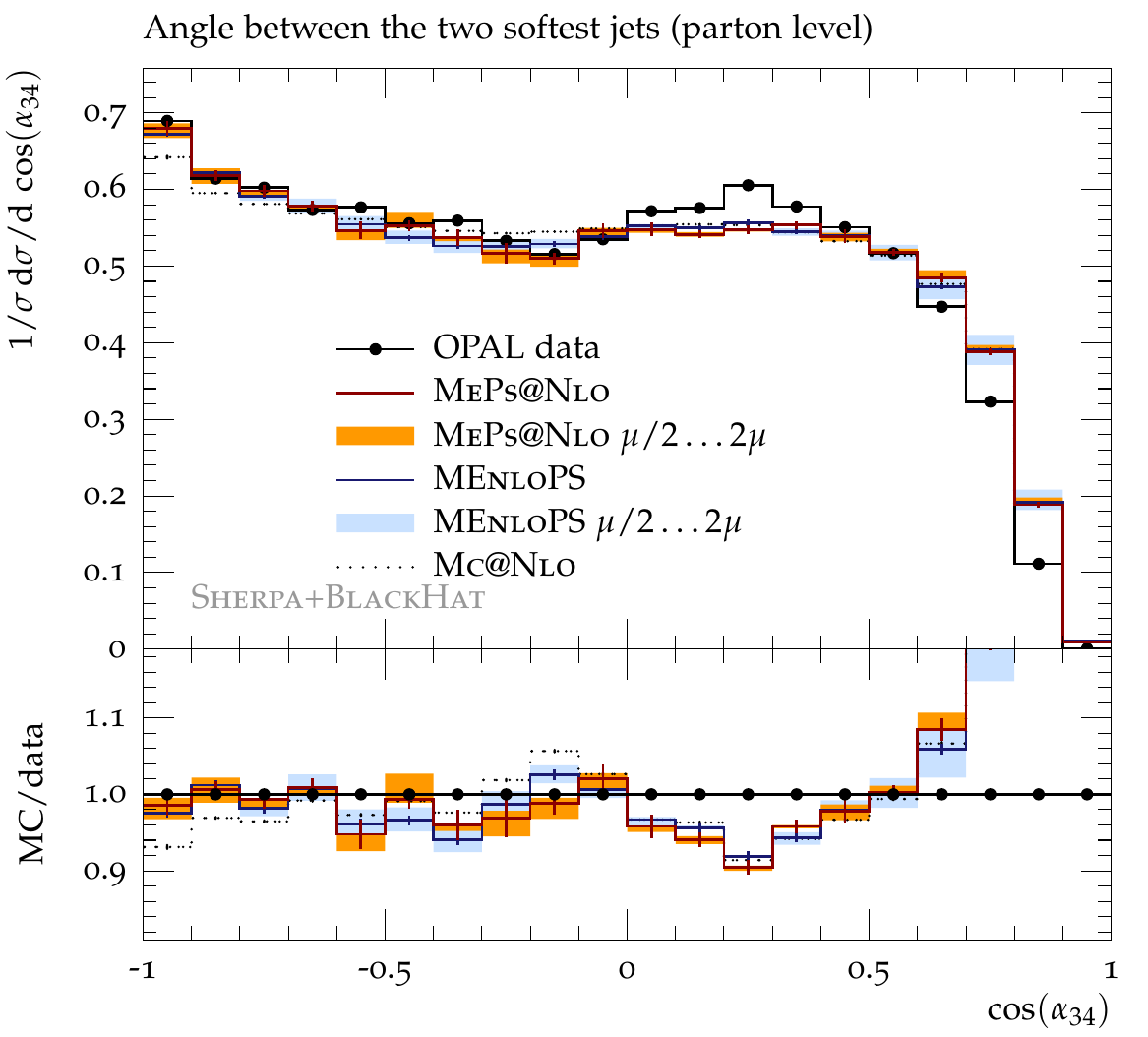}
  \end{center}
  \caption{Four-jet angles using the Durham algorithm compared to data from 
           OPAL~\cite{Abbiendi:2001qn}.}
  \label{fig:fourjetangles}
\end{figure}

In this section we compare the renormalisation scale uncertainties between the \MENLOPS and the \MEPSatNLO
method. We choose $\tilde{\mu}_R=C\mu_R$ with $C\in\{0.5,1,2\}$ and set $Y_{\rm cut}=2$.
In the \MEPSatNLO sample we
generate 2-,3-, and 4-parton final states at NLO and 5- and 6-parton final states at LO.
The \MENLOPS sample only has the 2-parton final state at NLO and the remaining multiplicities up to 6 partons from tree-level matrix elements.
Figures~\ref{fig:jetrates} to~\ref{fig:fourjetangles} show the respective scale variations
as bands around the central prediction with $C=1$. A significant reduction of the scale uncertainty 
is found for those observables, which are sensitive to the NLO parton-level results. This can be seen
in particular in Fig.~\ref{fig:jetrates}, where the $2\to3$ and $3\to4$-jet rates show significantly 
reduced uncertainties for larger $y$, while the $4\to5$ and $5\to6$-jet rates do not.
Similar effects are observed in most event shape distributions in the hard region, for example
in Fig.~\ref{fig:thrust}, for $T\to 0.5$.  The reduction of the scale 
uncertainty in the moments of the event shape distributions in particular is
more than impressive.
It is also worth pointing out that the typical Sudakov shoulder at $C=0.75$ 
in the $C$-parameter, which is notoriously difficult to describe in fixed-order
calculations, now shows a remarkably smooth behaviour due to the successful
interplay of the different multiplicity contributions. 

A final comment, concerning the evaluation of theory uncertainties by scale
variations is in order here.  Clearly, there are two sources of perturbative
uncertainties: the one analysed here, which stems from the matrix element.
It is thus susceptible to variations of the renormalisation and, if present,
the factorisation scale.  In addition, changes in the value of $\alpha_s$, 
which we did not pursue here, or in parton distribution functions would have
to be considered for a more complete assessment of such uncertainties.  On
the other hand, there are, of course, also uncertainties in the treatment of
secondary emissions through the parton shower.  There, in addition to the
variations outlined above, one could also vary the parton shower starting
scale, $\mu_Q$, which is equivalent to a variation of the corresponding 
resummation scale in analytical calculations.  Obviously in regions that are
dominated by the parton shower, such a variation would give a more sensible
estimate of theory uncertainties than a variation of the scales in the
matrix element, that we focused on here.  As an example for this, consider
the low-$p_\perp$ regime of the differential jet rates $y_{ij}$,
$-\log y_{ij}\to\infty$.  There the bands obtained from a scale variation
in the matrix element regime are suspiciously small, and it is clear that
a variation of the resummation scale would yield larger uncertainties.
Another important source of uncertainty is the model for parton to hadron fragmentation.
The same, obviously is true for the \MEPSatNLO and the \MENLOPS method, since
in the small-$y$ region both exhibit a comparable formal accuracy.  A careful analysis 
of such effects, however, is beyond the focus of this paper, which discusses
improvements of our ability to generate inclusive samples of events by
increasing the formal accuracy of the matrix element part of the simulation.
We therefore postpone this discussion to future work.

\begin{figure}
  \begin{center}
    \includegraphics[width=0.42\textwidth]{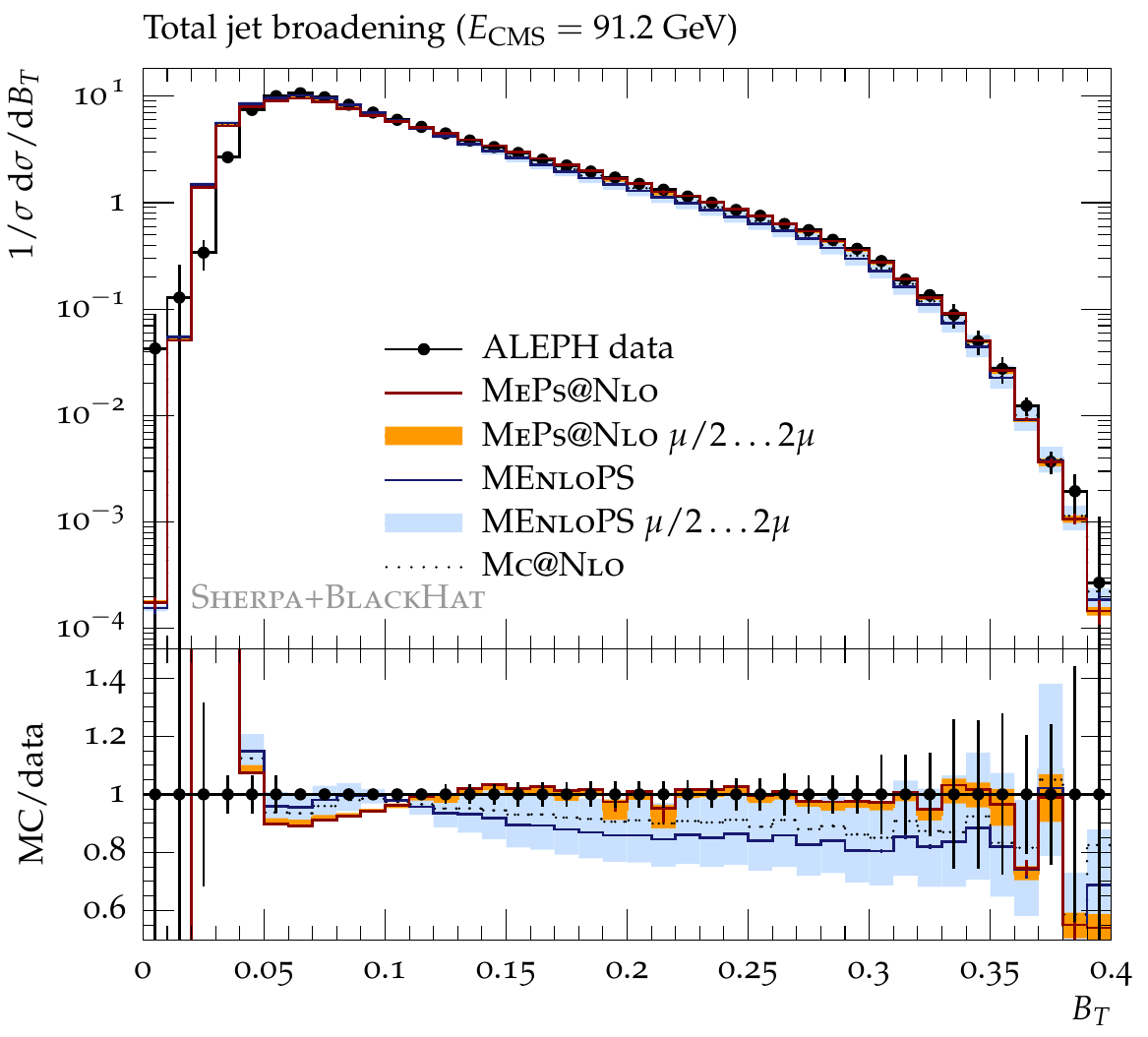}\nolinebreak
    \includegraphics[width=0.42\textwidth]{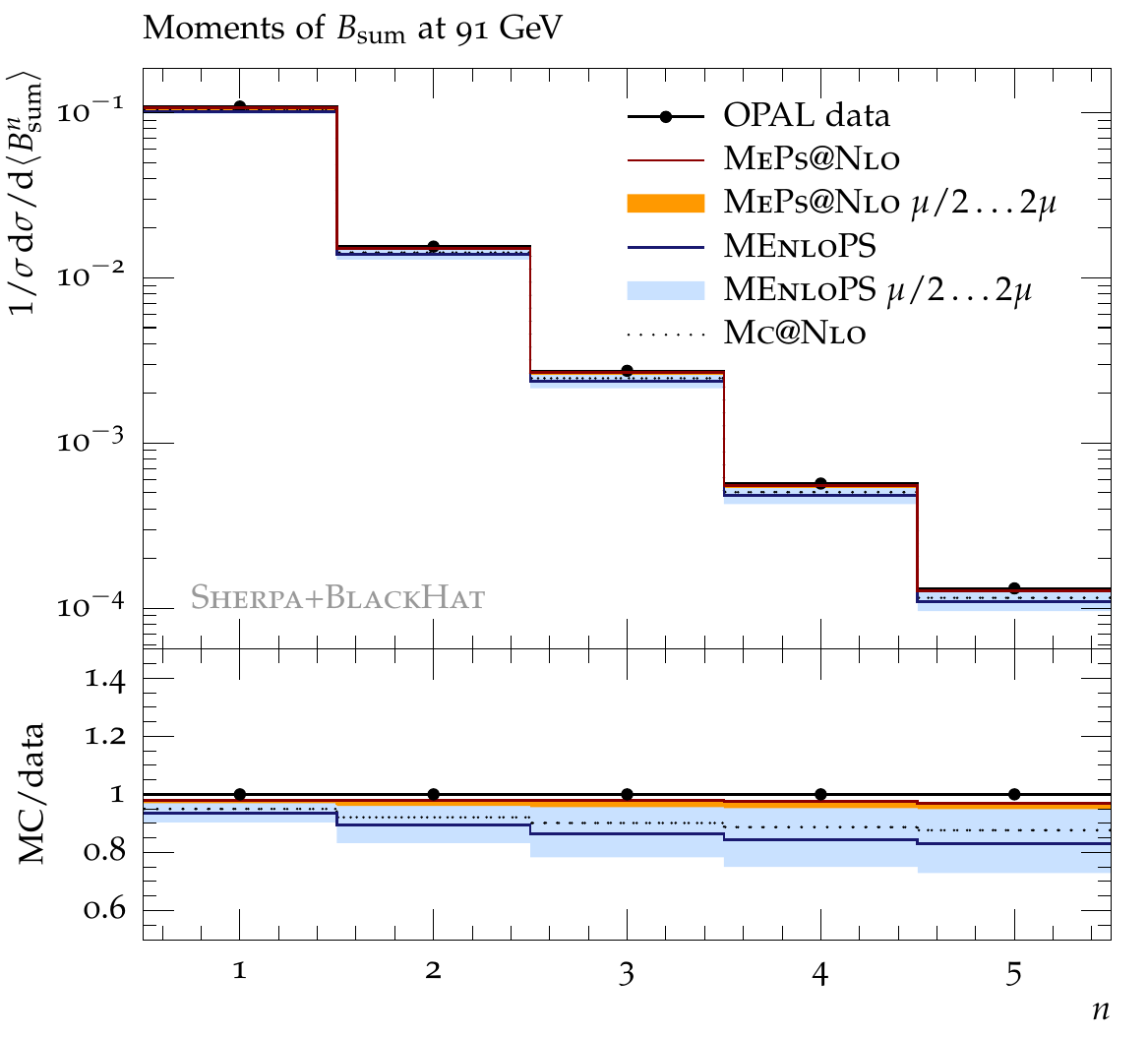}
  \end{center}
  \caption{Perturbative uncertainties in \protect\MENLOPS and \protect\MEPSatNLO 
           predictions of total jet/hemisphere broadening. Compared are the 
           measurements from ALEPH~\cite{Heister:2003aj} and OPAL~\cite{Abbiendi:2004qz}.}
\end{figure}

\begin{figure}
  \begin{center}
    \includegraphics[width=0.42\textwidth]{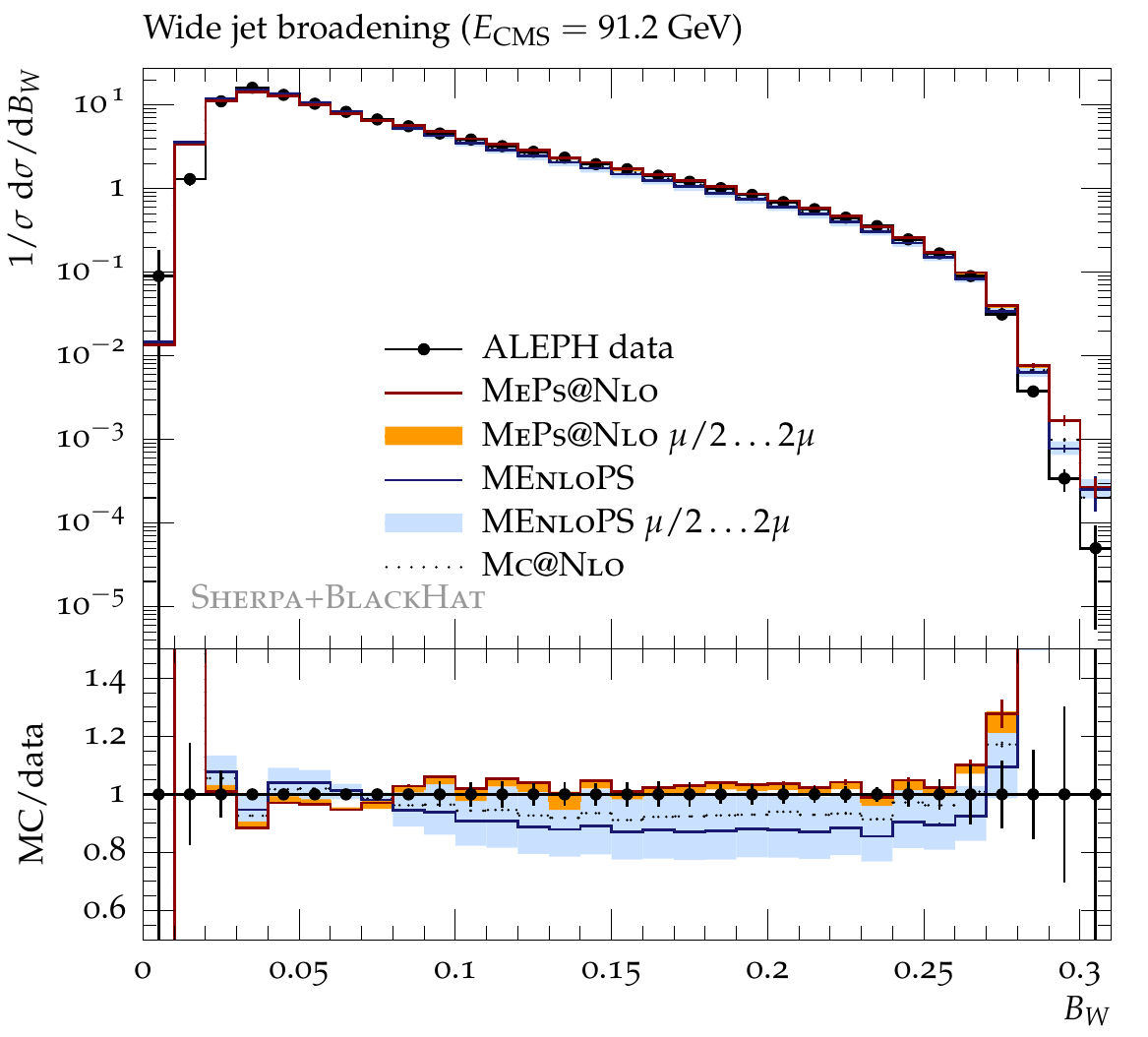}\nolinebreak
    \includegraphics[width=0.42\textwidth]{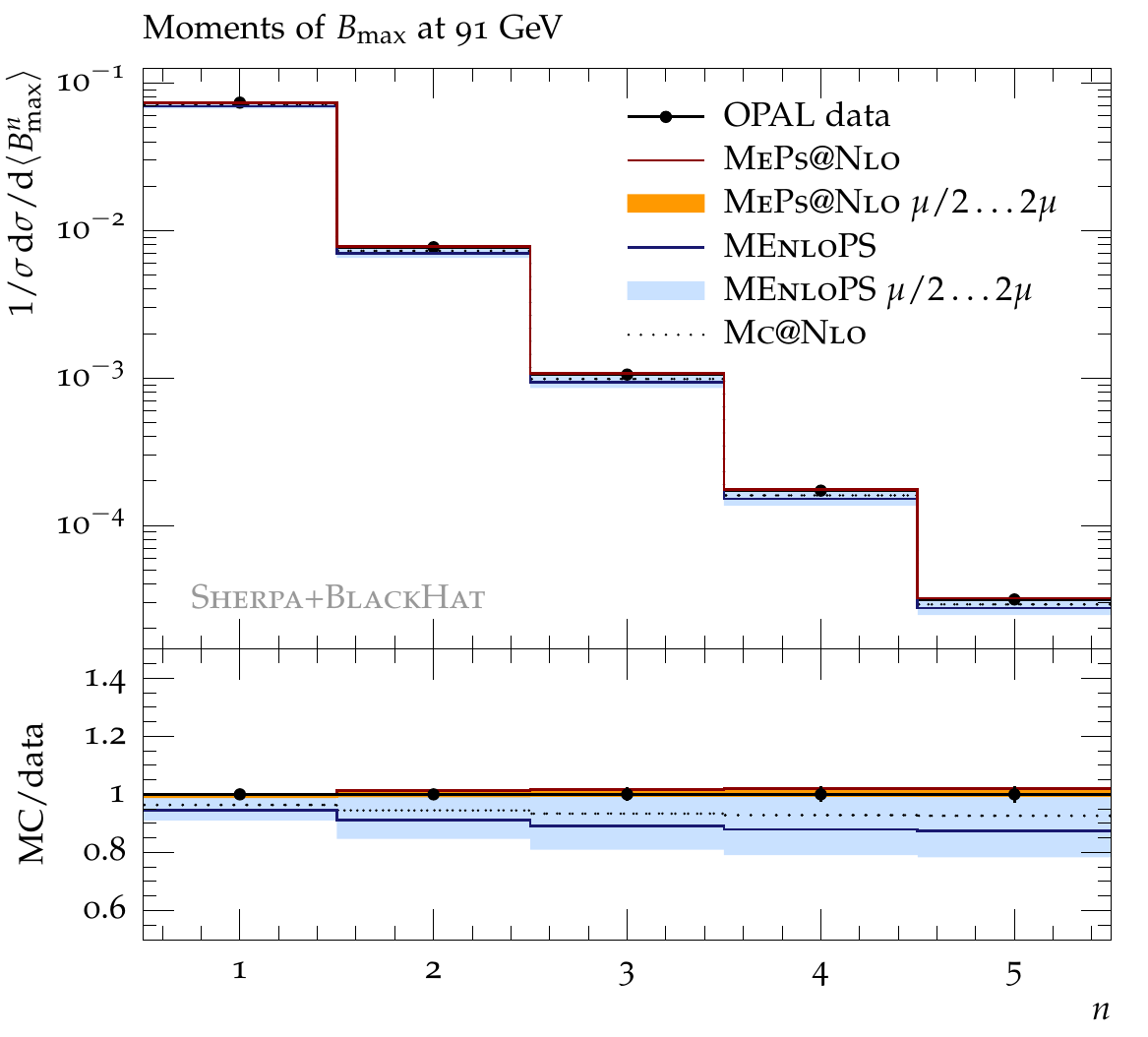}
  \end{center}
  \caption{Perturbative uncertainties in \protect\MENLOPS and \protect\MEPSatNLO 
           predictions of wide jet/hemisphere broadening. Compared are the 
           measurements from ALEPH~\cite{Heister:2003aj} and OPAL~\cite{Abbiendi:2004qz}.}
\end{figure}

\begin{figure}
  \begin{center}
    \includegraphics[width=0.42\textwidth]{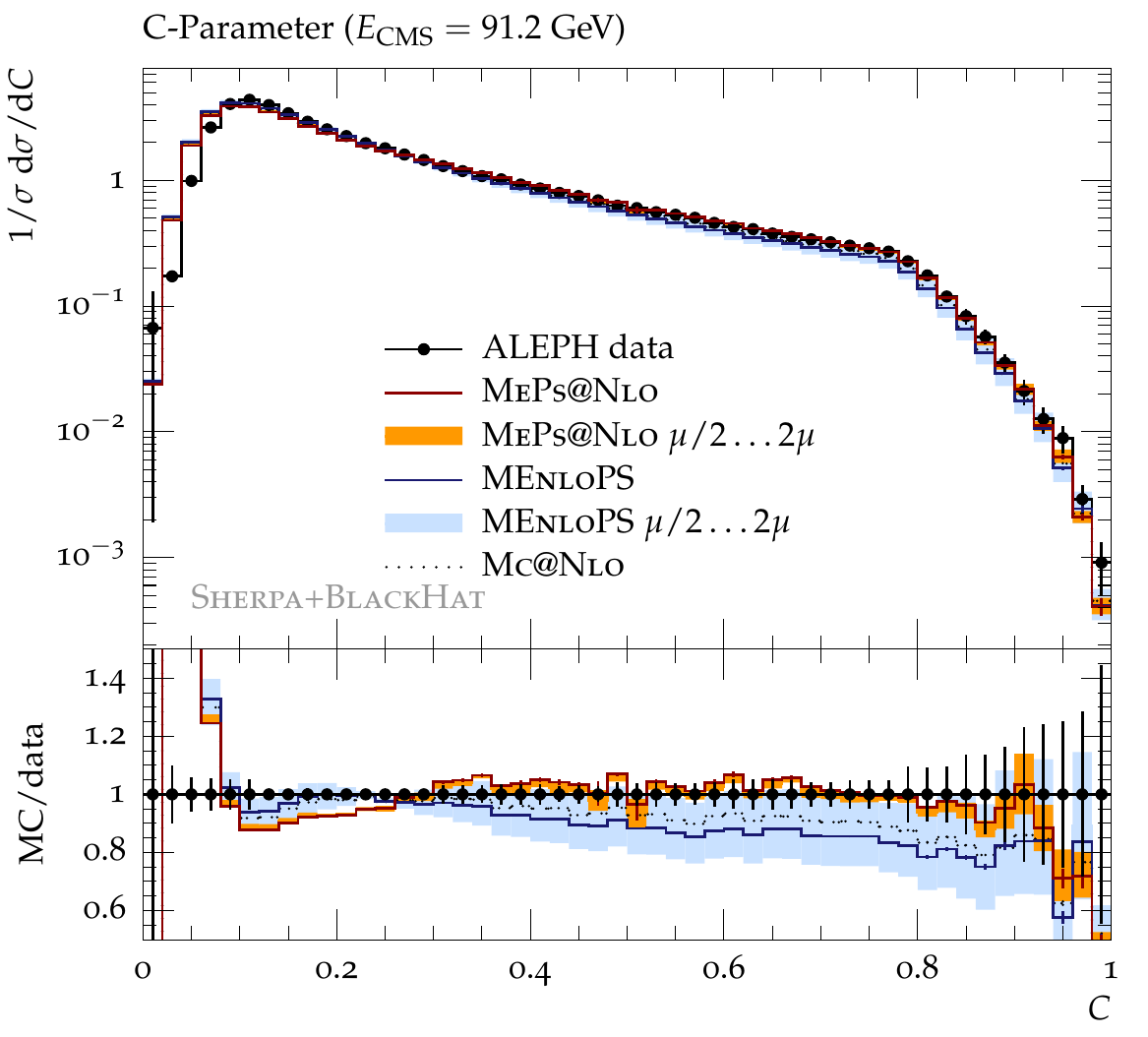}\nolinebreak
    \includegraphics[width=0.42\textwidth]{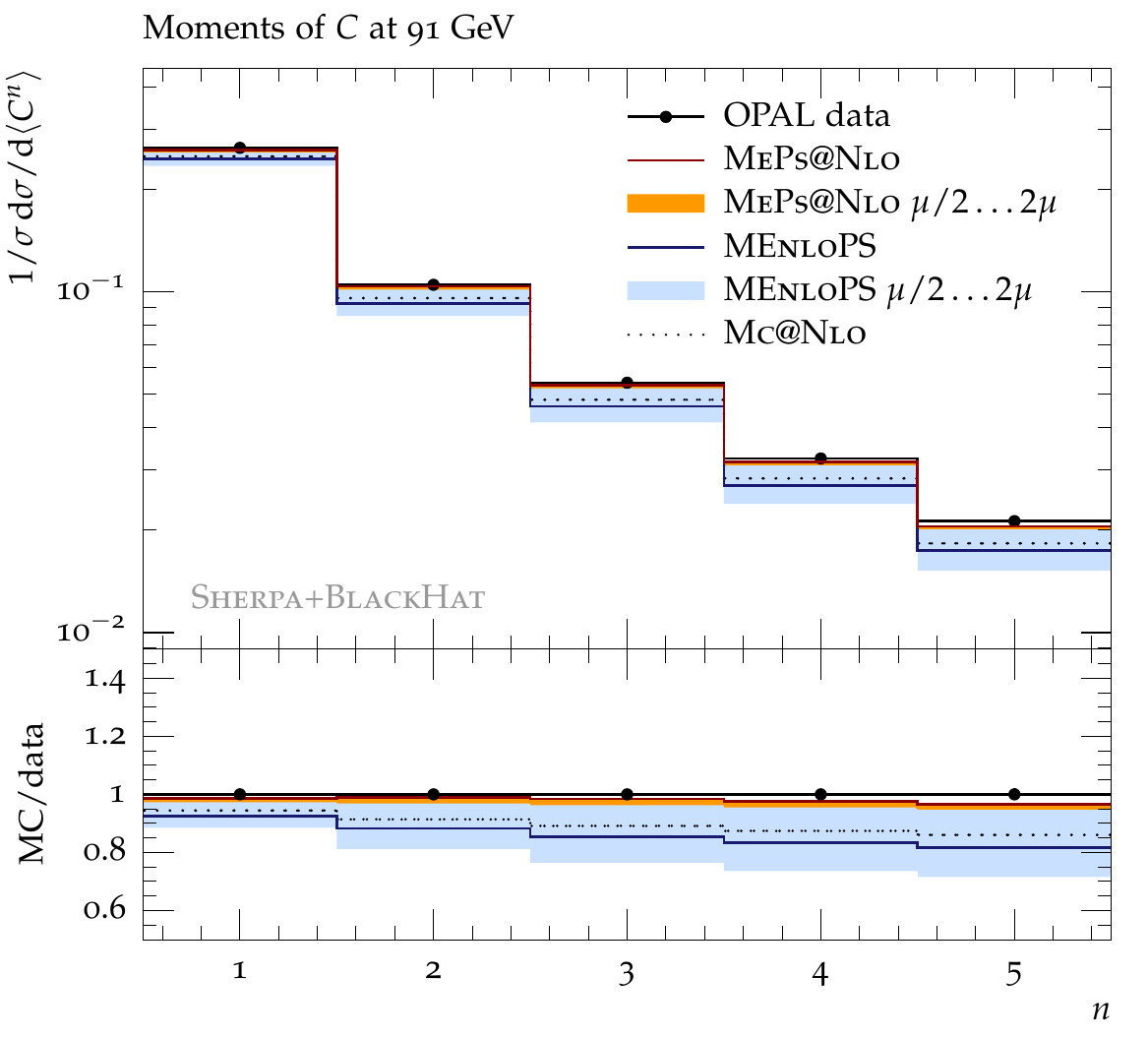}
  \end{center}
  \caption{Perturbative uncertainties in \protect\MENLOPS and \protect\MEPSatNLO 
           predictions of the $C$-parameter. Compared are the
           measurements from ALEPH~\cite{Heister:2003aj} and OPAL~\cite{Abbiendi:2004qz}.}
\end{figure}

\begin{figure}
  \begin{center}
    \includegraphics[width=0.42\textwidth]{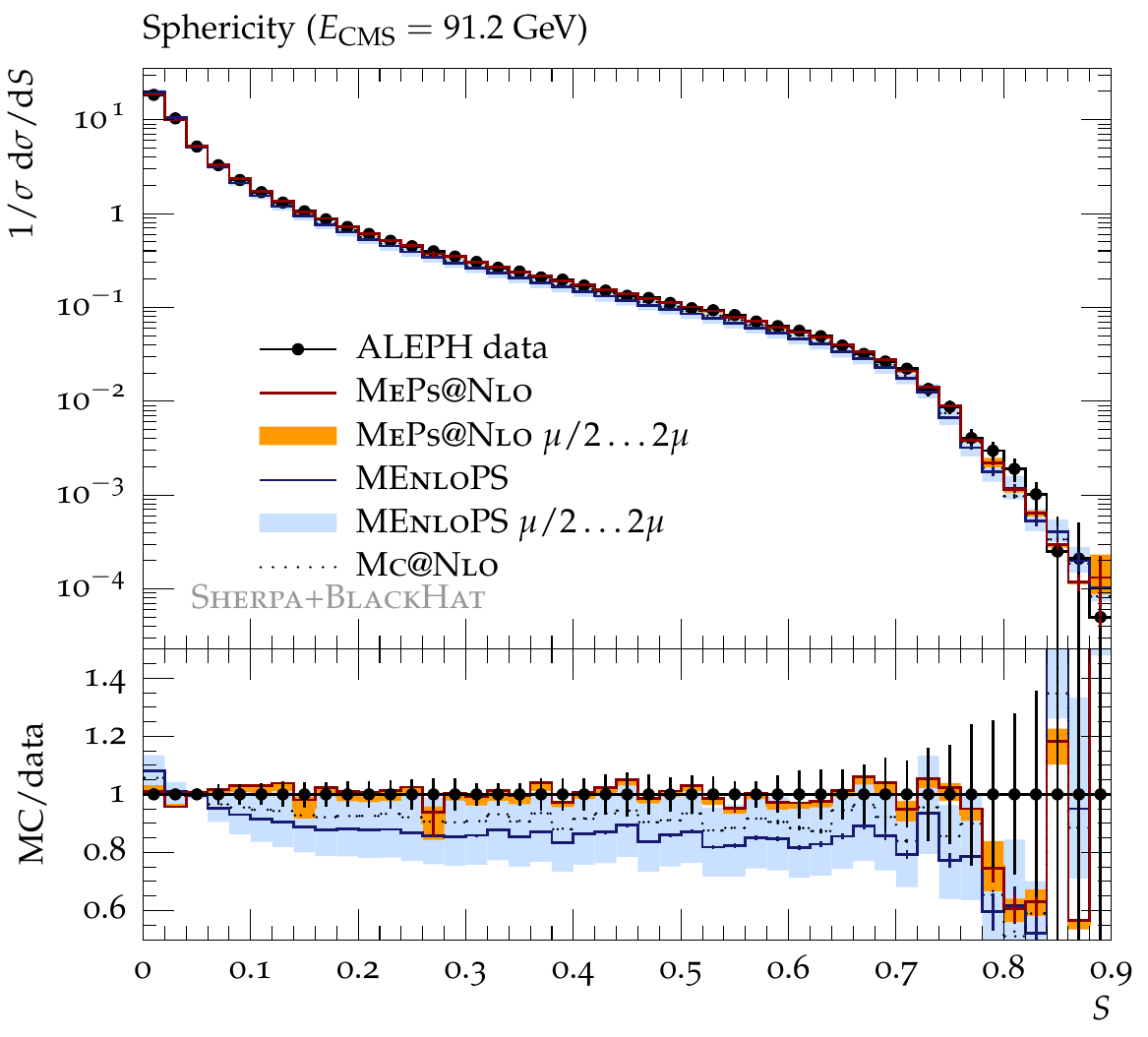}\nolinebreak
    \includegraphics[width=0.42\textwidth]{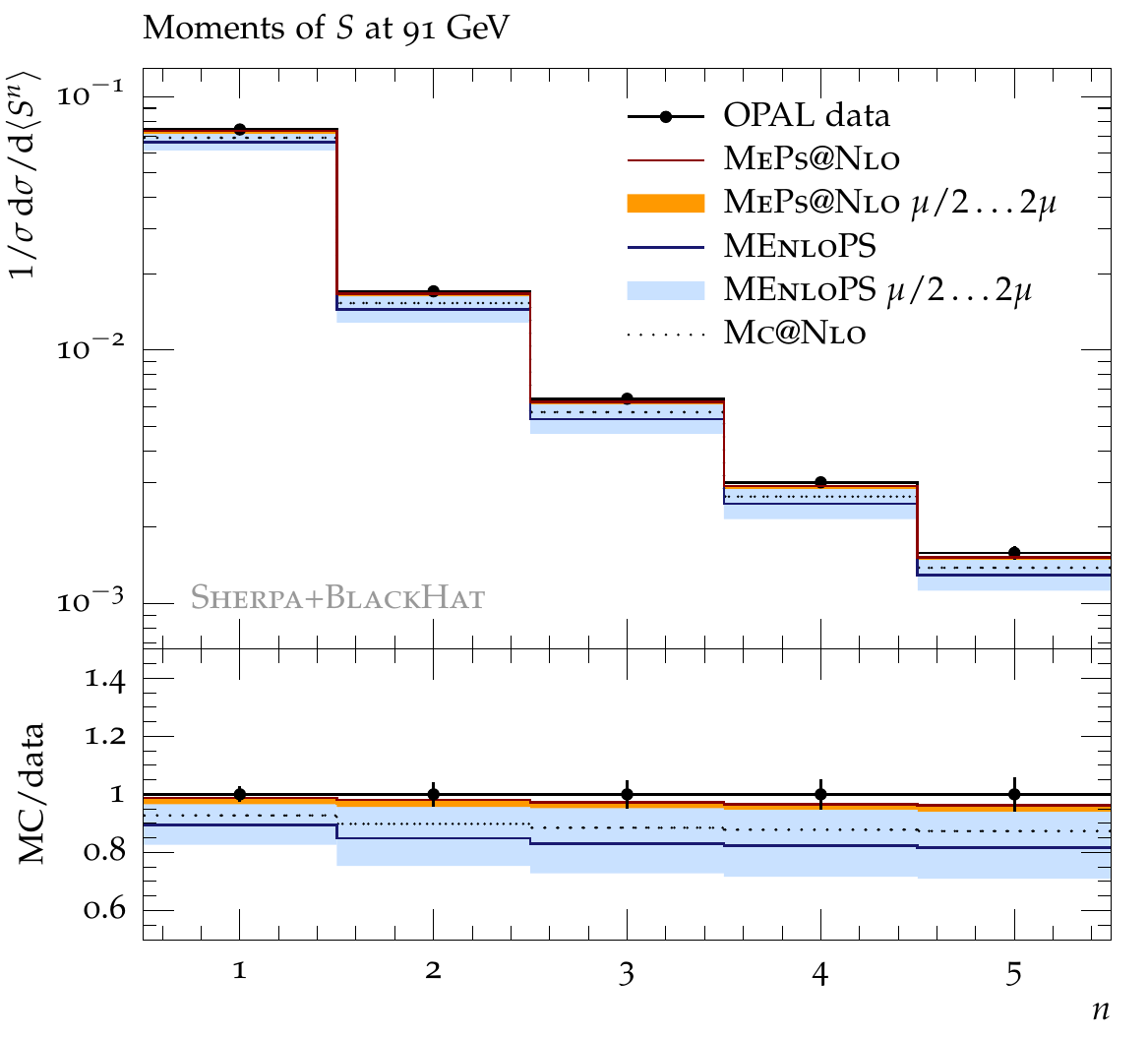}
  \end{center}
  \caption{Perturbative uncertainties in \protect\MENLOPS and \protect\MEPSatNLO 
           predictions of sphericity. Compared are the
           measurements from ALEPH~\cite{Heister:2003aj} and OPAL~\cite{Abbiendi:2004qz}.}
\end{figure}

\section{Conclusions}
\label{Sec:Conclusions}
In this paper we have introduced a method for a consistent multijet merging
at NLO accuracy for the case of $e^+e^-$-annihilations to hadrons.  By explicit
calculation, we have shown that our description maintains the higher order 
accuracy of the underlying matrix elements in their respective phase space
range, while the logarithmic accuracy of the parton shower is respected.  We 
have also analysed the impact of renormalisation scale variations in our new
formalism.  The results displayed here are exemplary for a far wider range of 
observables, which show a very good agreement between our simulation and data
throughout.  The most remarkable feature of our formalism is the greatly
reduced uncertainty due to variations of the renormalisation scale.  We have
also implemented our formalism for the case of collisions with hadronic initial
states \cite{Hoeche:2012yf}, where we find a similar behaviour.

\section*{Acknowledgements}

SH's work was supported by the US Department of Energy under contract 
DE--AC02--76SF00515, and in part by the US National Science Foundation, grant 
NSF--PHY--0705682, (The LHC Theory Initiative).  MS's work was supported by 
the Research Executive Agency (REA) of the European Union under the Grant 
Agreement number PITN-GA-2010-264564 (LHCPhenoNet). FS's work was supported
by the German Research Foundation (DFG) via grant DI 784/2-1.
This research is also supported in part by the Swiss National Science 
Foundation (SNF) under contracts 200020-138206, and by the Research Executive 
Agency (REA) of the European Union under the Grant Agreement number 
PITN-GA-2010-264564 (LHCPhenoNet).

TG and SH would like to acknowledge the Kavli Institute for Theoretical 
Physics (KITP) at UC Santa Barbara for its hospitality during the 2011 program  
''Harmony of Scattering Amplitudes".

We gratefully thank the bwGRiD project for computational resources.

\clearpage
\bibliographystyle{bib/amsunsrt_mod}  
\bibliography{bib/journal}
\end{document}